\documentclass{article}

\usepackage{aas_macros}

\usepackage[english]{babel}
\usepackage[section]{placeins}

\usepackage[letterpaper,top=2cm,bottom=2cm,left=3cm,right=3cm,marginparwidth=1.75cm]{geometry}

\usepackage{amsmath}
\usepackage{graphicx}
\usepackage[colorlinks=true, allcolors=blue]{hyperref}
\usepackage{authblk}

\usepackage{natbib}

\usepackage{longtable}

\newcommand{\Lfour}{L$_4$\,}
\newcommand{\Lfive}{L$_5$\,}

\date{August 16, 2022}

\begin{document}
\title{Implications for the collisional strength of Jupiter Trojans from the Eurybates family}

\author[1,2]{Raphael Marschall}
\author[1]{David Nesvorn\'y}
\author[1]{Rogerio Deienno}
\author[3]{Ian Wong}
\author[1]{Harold F. Levison}
\author[1]{William F. Bottke}

\affil[1]{Southwest Research Institute, 1050 Walnut St, Suite 300, Boulder, CO 80302, USA}
\affil[2]{CNRS, Observatoire de la Côte d'Azur, Laboratoire J.-L. Lagrange, CS 34229, 06304 Nice Cedex 4, France; raphael.marschall@oca.eu}
\affil[3]{Department of Earth, Atmospheric, and Planetary Sciences, Massachusetts Institute of Technology, Cambridge, MA 02139, USA}

\maketitle

\begin{abstract}
\noindent In this work, we model the collisional evolution of the Jupiter Trojans and determined under which conditions the Eurybates-Queta system survives.
We show that the collisional strength of the Jupiter Trojans and the age of the Eurybates family and by extension Queta are correlated.\\

\noindent The collis\textit{}ional grinding of the Jupiter Trojan population over $4.5$~Gy results in a size-frequency distribution (SFD) that remains largely unaltered at large sizes ($>10$~km) but is depleted at small sizes (10~m to 1~km).
This results in a turnover in the SFD, the location of which depends on the collisional strength of the material.
It is to be expected that the Trojan SFD bends between 1 and 10~km.\\

\noindent Based on the SFD of the Eurybates family, we find that the family was likely the result of a catastrophic impact onto a $100$~km rubble pile target.
This corresponds to objects with a rather low collisional strength \citep[10 times weaker than that of basaltic material studied in][]{BenzAsphaug1999}.\\

\noindent Assuming this weak strength, and an initial cumulative slope of the size frequency distribution of $2.1$ between diameters of 2~m and 100~km when the Trojans were captured, the existence of Queta, the satellite of Eurybates, implies an upper limit for the family age of 3.7~Gy.\\

\noindent Alternatively, we demonstrate that an unconventional collisional strength with a minimum at $20$~m is a plausible candidate to ensure the survival of Queta over the age of the Solar System.\\

\noindent Finally, we show how different collisional histories change the expected number of craters on the targets of the Lucy mission and that Lucy will be able to differentiate between them.

\end{abstract}

\section{Introduction} \label{sec:introduction}

The Lucy mission \citep{Levison2017LPSC} will be the first mission to study Jupiter Trojans (JT) up close. 
During five encounters for seven Jupiter Trojans between 2027 and 2033 it will sample the diversity of Trojans. 
The targets include (3548) Eurybates a C-type Trojan and the largest member of a collisional family \citep{BrozRozehnal2011}, (11351) Leucus, and (21900) Orus, two D-type Trojans, (15094) Polymele, a P-type Trojan, and (617) Patroclus and Menoetius, an almost equal mass binary pair that is also a P-type.
Polymele is the smallest target ($D\sim21$~km), whereas Patroclus is the largest ($D\sim140$~km). 
Leucus stands out for its very long rotation period \citep[$\sim445$~h;][]{Buie2018,Mottola2020}. 
Lucy has several objectives with the Trojans: (i) It will map the colors, albedos, and shapes of the target bodies; (ii) It will determine the distribution of minerals, ices and organics on each target surface; (iii) It will measure the crater size frequency distributions (SFD) on each target body to determine the relative ages of different parts of their surfaces,  and (iv) it will measure the masses and bulk densities of its targets as well as search for satellites and rings around them.

Jupiter Trojans are outer solar system planetesimals that orbit the Sun in a 1:1 resonance with Jupiter.
They are found in two swarms around the L$_4$ and L$_5$ Lagrangian points of the Sun-Jupiter system where they lead/trail Jupiter by $60^{\circ}$ respectively \citep{Lagrange1772}. 
Jupiter Trojans are known to be quite stable over the age of the solar system with only $\sim25\%$ having escaped the resonance since they were captured \citep{Levison1997,DiSisto2014,Holt2020}.
Currently we know of roughly 9,400 Jupiter Trojans \citep[$\sim3,000$ larger than 10~km;][]{Emery2015} making the population smaller than e.g. the main belt asteroid population \citep[$\sim10,000$ larger than 10~km;][]{Bottke2015}. 
Further, the \Lfour swarm, with $\sim1,800$ Trojans larger than 10~km, appears to have more objects than the \Lfive swarm, with $\sim1,200$ larger than 10~km \citep{Jewitt2000,YoshidaNakamura2008,YoshidaTerai2017}.

The JTs show a bi-modal color distribution \citep[``red" and ``less red"][]{Szabo2007,Roig2008,Emery2011,Wong2014,Jewitt2018}. 
While the ``red" Trojans overlap with ``red" Kuiper belt objects (KBOs) and Centaurs, the ``less-red" Trojan population does not have a clear analogue in the KBO population \citep[Fig.~1 in][]{WongBrown2016}.
There is no analogue for the ``very red" KBOs in the JTs.

The Eurybates family is a cluster of C-type fragments with similar orbits within the \Lfour swarm.  
Their members not only have inclinations that are tightly confined to $7.5^{\circ} \pm 0.5^{\circ}$, but they are also bluer than the ``less-red" JTs.
This makes them fairly exceptional within the Trojans.

Jupiter Trojans overall have very dark surfaces with visible albedos of $0.07\pm0.03${, according to data from the WISE infrared survey spacecraft \citep[][]{Grav2011}.} 
The large JTs are darker with albedos around $0.05$ while small JTs are on average ``brighter" with albedos around $0.1$ but with a wide scatter between $0.03$ and $0.2$ \citep[Fig. 7 in ][]{Grav2011}. 
JTs also have similar albedos and colors to comets \citep[][]{Fornasier2007,Fornasier2015}.
JTs populate the full stable region of around the Lagrange points, including high inclinations ($30^{\circ}$), but typically have low eccentricities ($<0.15$). The strong excitation in inclinations is an important constraint on the origin of Jupiter Trojans.

\begin{figure}[ht]
	\centering
	\includegraphics[width=\textwidth]{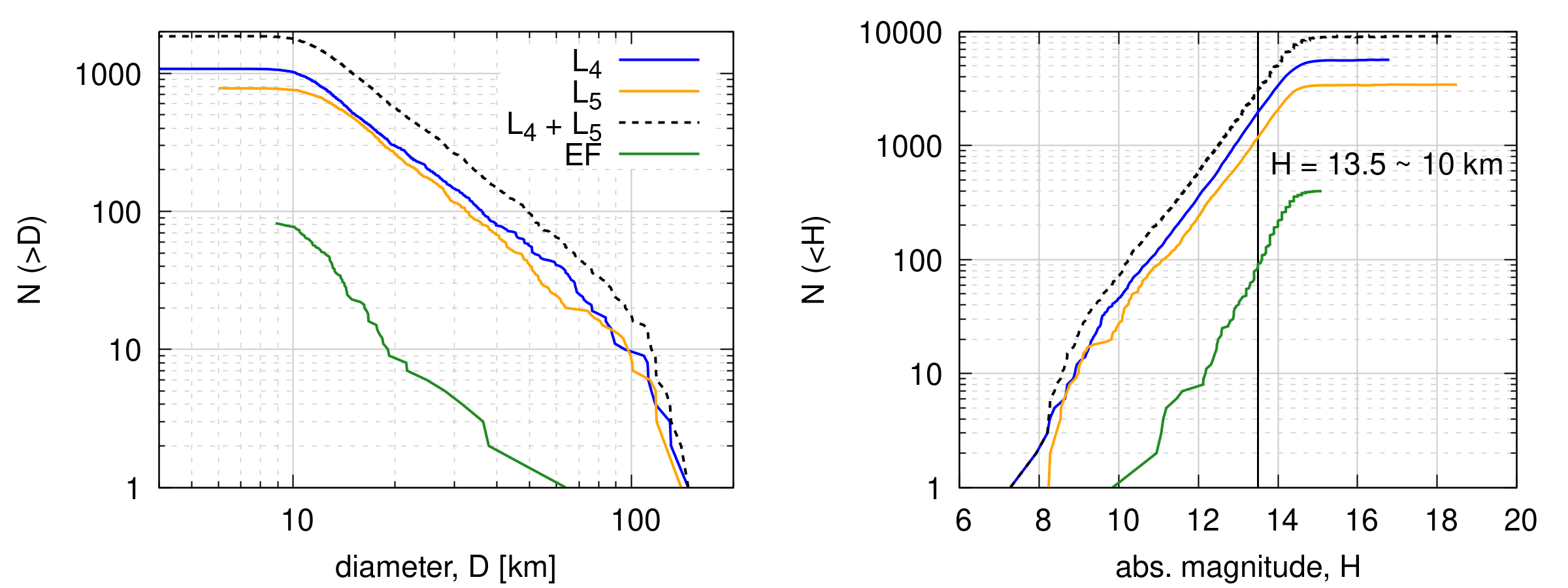}
	\caption{Left panel: The cumulative size-frequency distribution (SFD) as a function of diameter based on the WISE data \citep{Grav2011} of the \Lfour (blue line) and \Lfive (orange line) Jupiter Trojan swarms are shown. The dashed black line shows the combined Trojan SFD and the green line the one of the Eurybates family (EF). The WISE data is complete down to about $20$~km.
	Right panel: The cumulative SFD as a function of absolute magnitude, H, according to the data from the Minor Planet Center is shown. An absolute magnitude of 13.5 corresponds roughly to a 10~km Trojan, assuming an average albedo of 0.07 \citep{Grav2011}. The line colors correspond to the same (sub-) populations as in the left panel.}
	\label{fig:TrojanSFD} 
\end{figure}

Dynamical models suggest Jupiter Trojans are unlikely to have formed at their current location \citep[][]{Marzari2002}. 
This has led several groups to propose different origin scenarios and capture mechanisms.
For example, in the ``jumping Jupiter" model \citep{Gomes2005,Tsiganis2005,Morbidelli2005,Nesvorny2013ApJ} the primordial Kuiper belt beyond Neptune gets scattered by the giant planet instability. 
One part of the population is scattered outwards into the trans-Neptunian object (TNO) region which includes the Edgeworth-Kuiper belt.
Another part is scattered inwards.
It is there where Jupiter, that jumps in semi-major axis, can capture some of these objects in the 1:1 resonance.
This scenario would also imply that KBOs and Jupiter Trojans have a common origin but later evolved differently resulting in different colors, e.g. due to the different collisional \citep{WongBrown2016} and thermophysical environment.
The advantage of this model lies in the fact that it accurately predicts both the population size of the Jupiter Trojans and their orbital distribution, in particular their large range of inclinations.
In a second fairly recent model proposed by \cite{Pirani2019}, Jupiter forms at $\sim20$~au or farther and subsequently migrates to its current location while growing in mass and sweeping up planetesimals in the process and capturing them in the 1:1 resonance.
This model accounts for the asymmetry in the size of the two Trojan swarms (see Fig.~\ref{fig:TrojanSFD}), but the captured bodies do not reproduce the observed masses or the inclinations of the observed Trojans.

In this work we study the collisional evolution of the Jupiter Trojans and the Eurybates family to understand their implications for i) their material strength, and ii) the age of the Eurybates family and newly discovered satellite Queta \citep{Noll2020}. 
Queta's survival has implications for the family forming event.
We find that to a degree the age and strength of the Eurybates family and Queta are correlated and moreover depend on the SFD of the Jupiter Trojans at small sizes ($<1$~km). 

\newpage

\section{Constraints from the Trojans, Eurybates family, and Queta}\label{sec:constraints}

\noindent Three constraints are of particular importance for this work:
\begin{enumerate}
   \item the SFD of the Jupiter Trojans;
   \item the Eurybates family and its SFD; and
   \item the existence of Queta, the satellite of Eurybates.
\end{enumerate}
We will present these constraints in the order given above which also corresponds to going from a macro to the micro view.

\subsection{The Jupiter Trojan SFD}
First, Figure~\ref{fig:TrojanSFD} shows the cumulative SFD of the two Trojan swarms using the WISE data \citep{Grav2011}.
We often find that SFDs follows a power law where the number of objects larger than a certain diameter can be written as
\begin{equation}
    N(>D) = C_0D^{-q}; \quad C_0 = N_0 D_0^q;
\end{equation}
where $q$ is the cumulative power law exponent or slope, and $C_0$ is the normalization constant such that for a reference diameter, $D_0$, the number of objects larger than $D_0$ is $N_0$.
The differential power-law exponent, $p$, differs with respect to the cumulative power-law exponent by one: $p=q+1$.

The Jupiter Trojan SFD has a well-defined cumulative slope of $q=2.1$ between $10-100$~km for both the \Lfour and \Lfive swarms (Fig.~\ref{fig:TrojanSFD}).
At larger sizes the slope becomes steeper (similar to the KBO SFD) but is poorly defined because of the limited number of objects.
Note that 100~km planetesimals are thought to be the preferred size \citep[e.g.][]{Morbidelli2009b,KlahrSchreiber2020,KlahrSchreiber2021} for the formation of planetesimals by streaming instability \citep{YoudinGoodman2005}.
On the other end of the size range the population of small Trojans has been explored for \Lfour \citep{Jewitt2000,YoshidaNakamura2008,WongBrown2015,YoshidaTerai2017}.
The latter three studies find that the slope of the SFD continues below 10 km, below which there is some indications that the SFD might become shallower.
Unfortunately, the papers do not agree where (or if) this turnover takes place.
The latest study \citep{YoshidaTerai2017}, which uses the powerful Hyper Suprime-Cam attached to the Subaru Telescope, finds that a single power law is a better overall fit than a broken power law. 
It remains uncertain whether a bend in the Jupiter Trojan SFD has actually been detected.
If such a turnover exists it remains an open question how it relates to the observed turnover in the Edgeworth-Kuiper belt which occurs at roughly 1~km \citep{Singer2019,Morbidelli2021}.

\subsection{The Eurybates family}\label{sec:EurybatesFamily}
Second, we discuss the Eurybates family.
To identify the family we use the proper elements computed by Mira Bro\v{z} \citep{Holt2020} using the hierarchical clustering method  \citep[HCM,][]{Zappala1990,Nesvorny2015} with a cutoff of $40$~m/s. 
We have varied the cutoff to probe the change in family definition.
Increasing the cutoff leads to clustering in the wider neighborhood while a smaller cutoff leaves out many peripheral members. 
In total the HCM identified 400 family members which roughly doubles the family members previously found by \cite{Nesvorny2015}.
This change reflects the increased number of known JTs since the time of that paper.
The full list of family members with relevant properties presented in this work can be found in Tables~\ref{tab:EurybatesFamily} and \ref{tab:EurybatesFamilyColour}.

\begin{figure}[ht]
	\includegraphics[width=\textwidth]{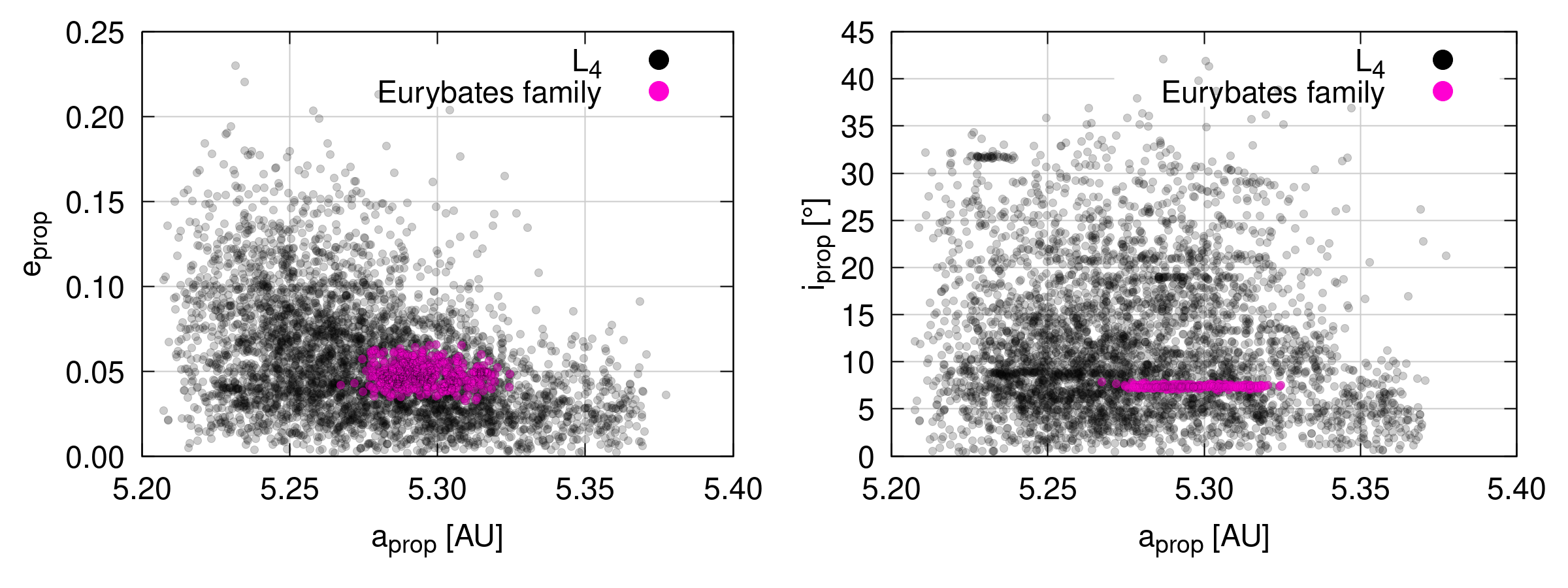}
	\caption{The proper elements of the \Lfour swarm (black circles) and the Eurybates family (pink circles) are shown. The left panel shows the proper semi-major axis, $a_{prop}$, and proper eccentricity, $e_{prop}$, while the right panel shows the proper inclination, $i_{prop}$. The family is particularly well defined in inclination and hence easily stands out in the right panel. }
	\label{fig:propEl} 
\end{figure}

The Eurybates family is well defined with respect to the proper elements (Fig.~\ref{fig:propEl}) and is roughly located in $a_{prop}\in [ 5.27$ au$, 5.32$ au$ ]$ , $e_{prop}\in [ 0.032, 0.066 ]$ , and $i_{prop}\in [ 6.99^{\circ}, 7.86^{\circ} ]$.

The family is located close to the edge of the stable orbit region \citep[Fig.~\ref{fig:propEl}, and ][]{Robutel2006}. 
In inclination the family is extremely well defined \citep[within half a degree for most members;][]{BrozRozehnal2011} and thus stands out in the right panel of Fig.~\ref{fig:propEl}.

\begin{figure}[hbt]
	\includegraphics[width=\textwidth]{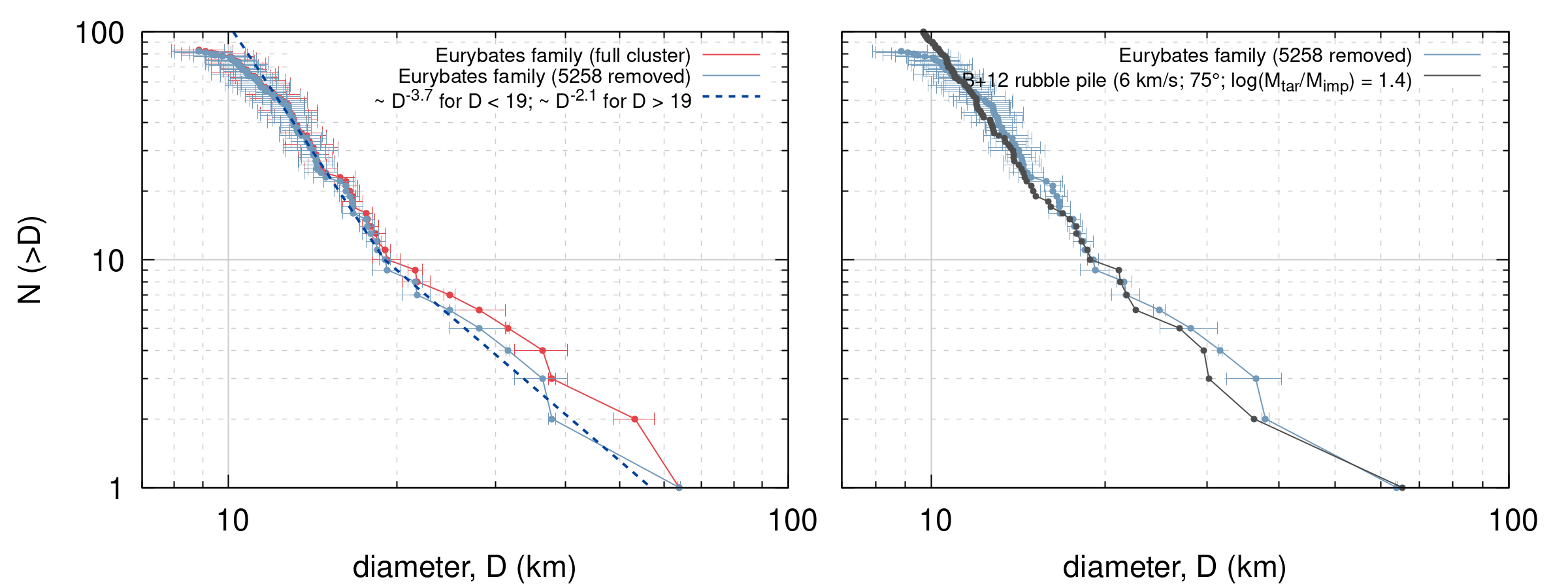}
	\caption{The left panel shows the SFDs of the Eurybates family with (i) all members retrieved by the HCM (red solid line and points) and (ii) suspected interloper (5258) Rhoeo removed from the family (blue solid line and points). The latter corresponds to our nominal case of the Eurybates family. The dashed line shows the approximate best power law fit for the nominal SFD with a slope of $2.1$ for objects larger than $19$~km and $3.7$ below that. The right panel shows a comparison of our nominal SFD with the results from the rubble pile simulation of \cite{Benavidez2012} at an impact speed of $6$~km/s, impact angle of $75^{\circ}$, and mass ratio of target to impactor bodies $\log(M_{tar}/M_{imp})=1.4$. The diameters and associated errors were taken from the NEOWISE data set v2.0 \citep[][]{Mainzer2019}.}
	\label{fig:EurybatesSFD} 
\end{figure}

Figure~\ref{fig:EurybatesSFD} shows the SFD of the Eurybates family as retrieved by the HCM (red line in left panel). 
This raw SFD has a shallow slope between $20-60$~km of $q\sim2.1$, similar to the overall JT slope (Fig.~\ref{fig:TrojanSFD}). 
At small sizes the SFD steepens up to a slope of $q\sim3.7$, a value that is steeper than that of the \Lfour Trojans.  
Slopes steeper than those of the background population are a common property of collisional families \citep{Durda2007}.

Before going any further, we need to establish if some of the identified members could be interlopers.
This is particularly important for the largest family member because the shape of the SFD at those sizes can be diagnostic of the family forming collision itself \citep[e.g.][]{Benavidez2012}.

To validate the possible detection of interlopers we first assess the likelihood of the background \Lfour population contaminating the Eurybates family region.
We have used the following dynamical constraint to determine the background population. 
First, we defined an ellipsoid in proper $(a,e,i)$-space to encircle the family.
The center of the ellipsoid was placed at $(a,e,i) = (5.30$ au$,0.049,7.4^{\circ})$ and the respective radii set to $(R_a,R_e,R_i) = (0.069$ au$,0.040,1.04^{\circ})$.
These values were chose to maximize the number of family members identified by HCM and minimizing the number of false defections.
All JTs within this ellipsoid were considered as a simplified definition of the family members and only slightly differed from the actual family members determined by HCM.
We then increased the volume of the ellipsoid by a factor of eight by increasing the radii by a factor of two in each dimension.
The Trojan density within this larger volume minus the family volume was used to estimate the number of interlopers we can expect in the family.
For objects with magnitude $H<12$ we expect on average $1.4$ interlopers. 
We also determined the probability of having $N$ interlopers. 
There is a $40\%$ chance that $N\ge2$, and a $20\%$ chance that $N\ge3$.
Consequently there is a non-negligible probability that there could be at least three large interlopers in the family.
Over all sizes about $8\%$ of the family members might be interlopers.

\begin{figure}[!ht]
	\centering
	\includegraphics[width=0.55\textwidth]{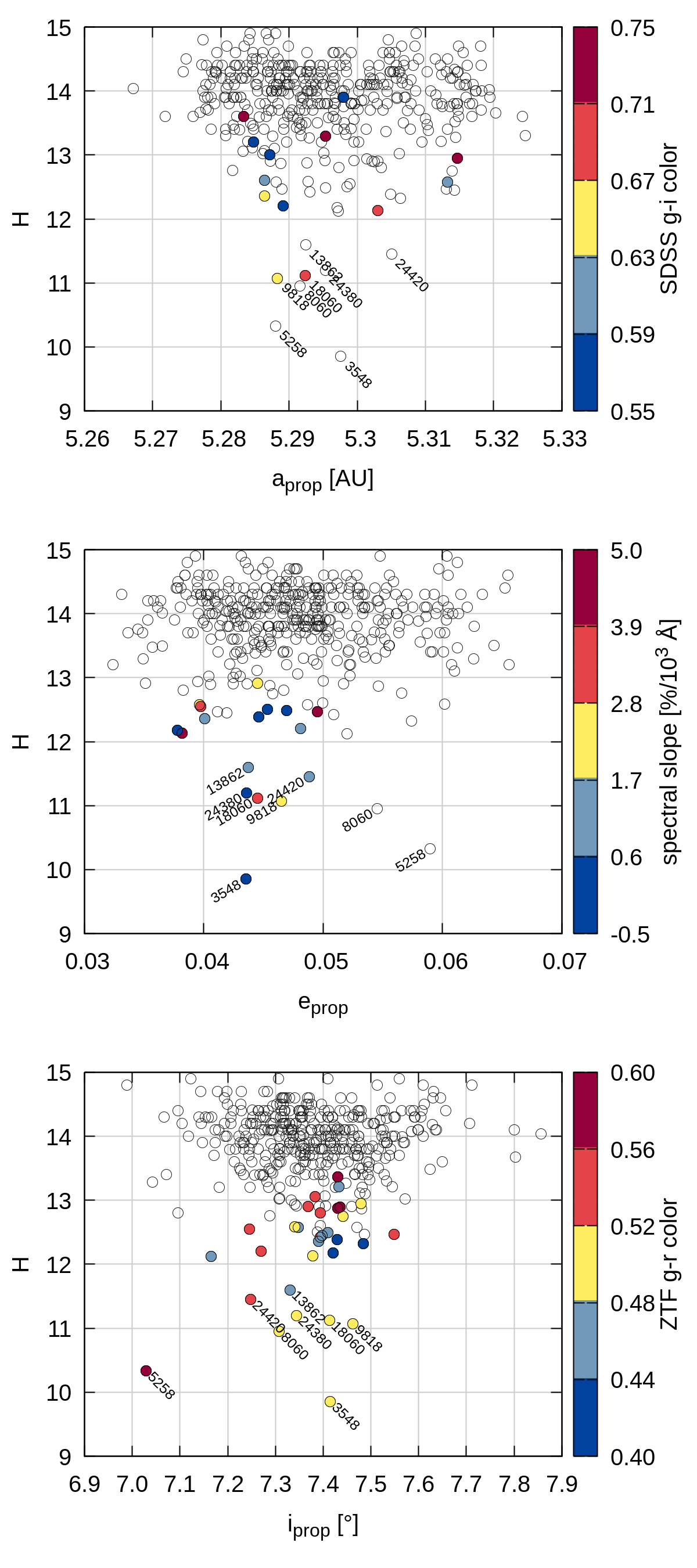}
	\caption{\small The top panel shows the H-magnitude of the Eurybates family as a function of proper semi-major axis as well as g-i color from the Sloan Digital Sky Survey, SDSS \citep{Ivezic2001}. The text labels indicate the Trojan numbers of the eight brightest family members. The middle panel shows the H-magnitude as a function of proper eccentricity and spectral slope by \cite{Fornasier2007}. For direct comparison the color scale has been chosen such that it corresponds roughly to the SDSS color bar in the top panel. The bottom shows the H-magnitude as a function of proper inclination and g-r color from the Zwicky Transient Facility Observations, ZTF \citep{SchemelBrown2021}. Trojan (5258) Rhoeo is a clear outlier with respect to proper eccentricity and inclination as well as ZTF color and therefore a very likely interloper. See Table~\ref{tab:EurybatesFamily} and \ref{tab:EurybatesFamilyColour} for all details on the family members depicted here.}
	\label{fig:FamilyColors} 
\end{figure}

To investigate which of the largest members might be interlopers, we inspected their proper orbital elements as well as color data (Fig.~\ref{fig:FamilyColors}).

We find that the second brightest/largest object identified by HCM, (5258) Rhoeo, is clearly displaced in proper inclination and eccentricity with respect to the center of the family. 
This would imply, that a very large fragment from the family forming event would have been ejected at a high speed. 
That is not to be expected. 
Collisonal families typically have their largest members in center of the orbital element distribution \citep[e.g.][]{Milani2017}.
That is a consequence of the velocity distribution being a function of $1/D$. 
We hence strongly suspect that (5258) Rhoeo is an interloper. 
This conclusion is strengthened by its anomalous color compared to the other family members.

We collected colors from the Sloan Digital Sky Survey, SDSS \citep{Ivezic2001}, spectral slopes between $0.3$ to $0.9~\mu$m \citep{Fornasier2007}, and colors from the Zwicky Transient Facility Observations, ZTF \citep{SchemelBrown2021}.
The SDSS g-i color (top panel in Fig.~\ref{fig:FamilyColors}) and the spectral slopes from \cite{Fornasier2007} (middle panel in Fig.~\ref{fig:FamilyColors}) are connected and thus their color scales have been chosen such that they correspond to each other.
The color scale of the ZTF data cannot be directly compared to the other two.

Eurybates and the bulk of the family have ZTF g-r of $\sim0.5$  whereas (5258) Rhoeo has $0.6$ which is significantly redder. 
This together with the orbital displacement of (5258) Rhoeo previously discussed provides strong evidence that it is indeed an interloper.

\begin{figure}[ht]
	\centering
	\includegraphics[width=\textwidth]{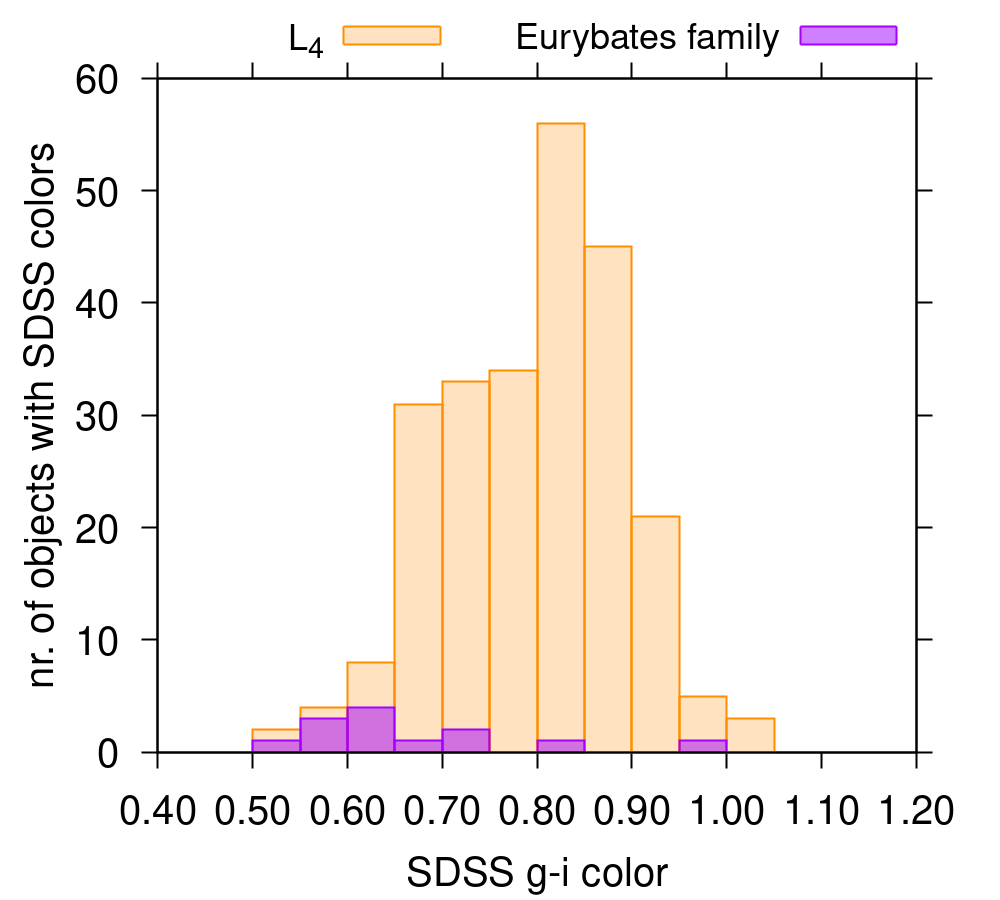}
	\caption{The histogram shows all L$_4$ Trojans (orange) and all Eurybates family members (purple) with known SDSS colors. The Eurybates family is significantly bluer than the L$_4$ Trojans overall.} 
	\label{fig:L4-vs-Eurybates-colors} 
\end{figure}

The Eurybates family is significantly bluer than the overall \Lfour population (Fig.~\ref{fig:L4-vs-Eurybates-colors}).
With a g-i color of $0.96 \pm 0.09$ (313024) 2000 AV210 is on the very red end of the color distribution and thus another likely interloper.
It has a diameter of $10.805 \pm 0.599$ and will thus not affect the family SFD and the work described here.

There are two other family members that might be interlopers. 
(8060) Anius has a somewhat larger eccentricity (middle panel in Fig.~\ref{fig:FamilyColors}) but its ZTF color is close to the average color of the family.
It is therefore unclear if (8060) Anius is indeed an interloper.  
We leave this as an open question.
Future survey data might be able to provide additional insights.
Similarly, (9818) Eurymachos has a slightly low semi-major axis for its size, comparable to (5258) Rhoeo (see top panel in Fig.~\ref{fig:FamilyColors}).
Other than that, the body is unremarkable with respect to the colors or orbital elements of the other family members.
For this reason, we do not consider (9818) Eurymachos an interloper in this work.

The other larger family members do not have unusual colors or orbital elements.
Consequently, for our work here we consider (5258) Rhoeo an interloper and exclude it from what we consider to be the nominal family.
The family SFD excluding (5258) Rhoeo is shown in Fig.~\ref{fig:EurybatesSFD}. 
We can use these data to explore what the SFD tells us about the family forming event.

A substantial amount of work has been done to simulate the outcome of asteroid impacts \citep[e.g.][]{BenzAsphaug1999,Durda2007,Benavidez2012,Jutzi2015,Benavidez2018,Jutzi2019} using primarily a Smoothed Particle Hydrodynamics, SPH, approach.
We have access to the simulation results from \cite{Durda2007}, who used 100 km diameter monolithic target bodies, and \cite{Benavidez2012}, covering collisions of 100 km rubble pile target bodies.
Given that these two works cover a large part of parameter space, we consider this to be a good first order reference for possible outcomes.

We used three criteria to determine whether a simulation outcome matched the observed family SFD: 1) the mass ratio of the two largest family members shall be within $25\%$, 2) the slope between the second and tenth largest object shall be within $\pm0.2$, and 3) the slope between the tenth and $30^{th}$ largest body shall be within $\pm0.2$.
These criteria were chosen because they fully characterize the observed family SFD and should so be fairly diagnostic of the family forming collision.
For the $387$ SPH runs available to us, the mass ratio varied between $3\times 10^{-5}$ and $1$ (median of $0.03$), the slope between the second and tenth largest object varied between $-0.6$ and $-15$ (median of $-3.7$), and the slope between the tenth and $30^{th}$ largest body varied between $-1.6$ and $-10$ (median of $-4.4$).

We found that only one model outcome out of 387 satisfied all three criteria (see right panel of Fig.~\ref{fig:EurybatesSFD}).
The model SFD is remarkably close to the actual family SFD.
The simulation corresponded to a collision between a $100$~km rubble pile target\footnote{Sometimes impact simulation results are scaled to different target sizes to fit family SFDs \citep[e.g.][]{Durda2004} but this was not needed here.} \citep{Benavidez2012} and a $\sim35$~km impactor ($\log(M_{tar}/M_{imp})=1.4$) at $6$~km/s and an impact angle of $75^{\circ}$ from vertical.
Though this match is unlikely to be a unique solution, it is an important piece of information to be evaluated.
This is the first indication in our analysis that Jupiter Trojans might be collisionally weak.  We will examine this issue below.

A strong constraint for our modeling would be the estimated age of the family, but unfortunately, it is  rather poorly constrained.
\cite{BrozRozehnal2011} and \cite{Rozehnal2016} estimated the family age between 1 and 4~Gy.
\cite{Milani2017} was unable to retrieve an age for the Eurybates family from their dynamical methodology.
Finally, using the escape probability of family members as derived from dynamical considerations, \cite{Holt2020} estimated the minimum age at $\ge 1.0 \pm 0.4$~Gy.
We are hence left at a weak constraint for the family age between 1~Gy and the age of the Solar System.

\subsection{Eurybates' satellite Queta}
Our last and arguably most important constraint comes from Eurybates itself.
\cite{Noll2020} reported the discovery of a $1.2\pm0.4$~km satellite around (3548) Eurybates, which has since been named Queta.
Eurybates is only the fourth Trojan with a known satellite \citep{Noll2020}.

Since its discovery, Queta's orbit has been further constrained by Hubble Space Telescope observations \citep{Brown2021}.
It has a semi-major axis of $2350\pm11$ km, an orbital period of $82.46 \pm 0.06$~days, and a small eccentricity of $0.125 \pm 0.009$ \citep{Brown2021}.
We find the low eccentricity to be particularly noteworthy because impact simulations \citep[e.g.][]{Durda2007} typically generate companions around the target body that have highly eccentric orbits. 
In dynamics, ``particles always return to the scene of the perturbation'' unless some mechanism has changed their orbit.
In Questa's case this can be due to the Kozai effect which lets Queta's orbit go through oscillations in eccentricity \citep[][]{Brown2021}.

The existence of Queta as a satellite of Eurybates, the most sizeable member of the largest collisional family within the Jupiter Trojans, immediately poses the question as to the long term survival of Queta.
To date, the dynamical stability of Queta has been studied in \cite{Brown2021}.  Here we will examine whether Queta can survive the expected collisional environment within \Lfour.
We will summarize the dynamical findings from \cite{Brown2021} in Sec.~\ref{sec:methods}.

\section{Methods}\label{sec:methods}
To simulate the collisional evolution of the Trojans we employ the \textit{Boulder} collisional code \citep{Morbidelli2009,Nesvorny2011AJ,Nesvorny2018NatAs}.
Here we only outline the broad principles of the code but a detailed description is provided in \cite{Morbidelli2009} and \cite{Nesvorny2018NatAs}.
\textit{Boulder} requires an initial SFD of the population, the intrinsic impact probability between bodies ($P_i$), average impact speeds between bodies, and a function that describes the critical impact energy $Q^*_D$,  defined as the energy per unit target mass needed to disrupt and disperse $50\%$ of the target \citep[e.g.][]{BenzAsphaug1999,Davis2002}.
The SFD and $P_i$ are used to calculate the expected number of impacts per time step of the impactor population of size $R_i$ on the targets with size $R_j$ (where $R_i<R_j$).
For a given impact \textit{Boulder} calculates the specific impact energy $Q$, which is defined as the kinetic energy of the impactor divided by the target mass.
Collisions where $Q < Q^*_D$ are referred to cratering events while collisions with $Q > Q^*_D$ correspond to super-catastrophic disruption events.
When a catastrophic disruption occurs, the code calculates the masses of the largest remnant and that of the largest fragment as well as the power-law slope of the smaller fragments.
These values are calculated based on the scaling laws found by the hydrocode results of \cite{Durda2004}, \cite{Durda2007}, and \cite{Nesvorny2006Icarus}.
Each collision alters the SFD which is then used to estimate the subsequent number of expected impacts.

Two fundamental properties of the JTs are unknown: 1) the SFD (in particular for $< 10$~km) when the Trojans were captured in resonance with Jupiter (we will call this the initial SFD), and 2) the collisional strength of Trojans as defined by $Q^*_D$.
With regards to the initial SFD, our tests suggest collisional evolution will not meaningfully affect the SFD of $D  \ge 10$~km Trojans. 
On this basis, we assume the shape of the Jupiter Trojan SFD for this size range is likely primordial and hence close to the current day SFD.
This will also become apparent from our results below.
On the other hand the shape of the SFD below 10~km is unknown for the initial population.
The present day SFD is complete to $\sim20$~km (see Fig.~\ref{fig:TrojanSFD}) and the slope is arguably well known down to $\sim3-10$~km \citep[][]{YoshidaNakamura2008,WongBrown2016,YoshidaTerai2017}.
We are therefore required to make assumptions about the initial SFD below 10~km. 
Here we will assume two end member cases. 
For the first, we will assume that the initial slope below 10~km is simply a continuation of the current day slope above 10~km (2.1, see Fig.~\ref{fig:TrojanSFD}).
For the second, we will assume that the initial slope below 10~km is one ($q=1$) and thereby significantly shallower than the slope between 10-100~km ($q=2.1$).
This signifies an initial depletion of the JT population at small sizes.


Our version of \textit{Boulder} also allows us to follow the orbital evolution of the Eurybates-Queta system in terms of how impacts on either Eurybates and Queta can alter the orbit of Queta or potentially disrupt either of them \citep{Nesvorny2011AJ,Nesvorny2018NatAs}. 
Changes to the semi-major axis and eccentricity of Queta are tracked over time.
For non-disruptive impacts on either of the binary components, \textit{Boulder} computes the change of the binary orbit depending on the linear momentum of the impactor as described in detail in the methods section of \cite{Nesvorny2018NatAs}.

We are particularly interested in quantifying the four ways the Eurybates-Queta system can be dissolved: 1) impacts on Eurybates that a) catastrophically disrupt Eurybates or b) displace Eurybates such that Queta is no longer in orbit, and 2) impacts on Queta that a) catastrophically disrupt Queta or b) kick Queta out of orbit.
Here we assume a semi-major axis of $2400$~km and an eccentricity of $0.1$, but we note that the orbit can evolve purely dynamically \citep{Brown2021}. 
In particular the eccentricity is predicted to oscillate between 0.1 and 0.35 with a roughly 500 year period.
This is induced when Queta's argument of pericenter sweeps by the Kozai resonance islands at $90^{\circ}$ and $270^{\circ}$ \citep{Brown2021}.
We have tested the sensitivity of our results by varying the semi-major axis between $1200$~km and $7200$~km, and the eccentricity between $0.1$ and $0.5$.
We have not found any significant change of our results within these ranges. 
For this reason, we will only present the nominal case.
But we will return to this point and see why in this case the survival of Queta is not sensitive to the binary semi-major axis.

\begin{figure}[ht]
      \centering
      \includegraphics[width=\textwidth]{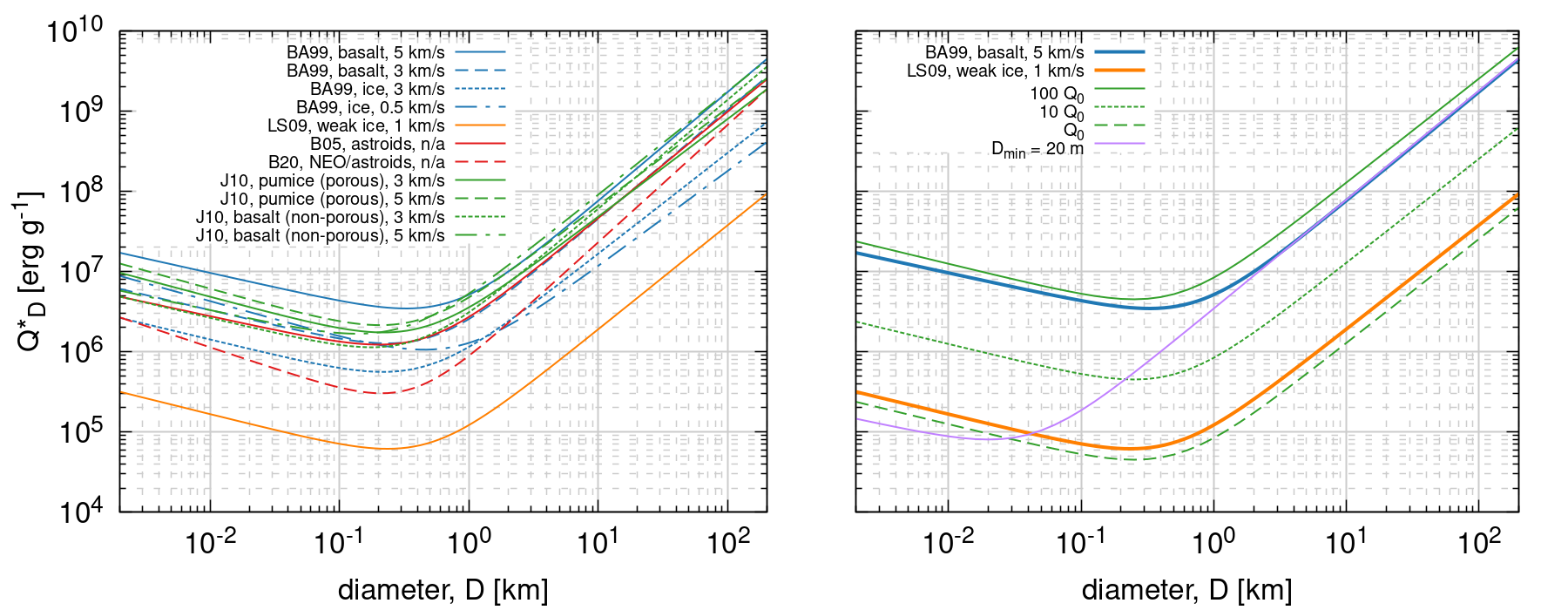}
      \caption{Critical impact energy, $Q^*_D$, according to different studies (BA99 \cite{BenzAsphaug1999}; LS09 \cite{LeinhardtStewart2009}; B05 \cite{Bottke2005}; B20 \cite{Bottke2020}; J10 \cite{Jutzi2010}) are shown in the left panel. The right panel shows three different model assumptions $Q_0$ (dashed green), $10 Q_0$ (dotted green), and $100 Q_0$ (solid green) as well as a $Q^*_D$ with a minimum at 20~m (purple line). In addition the literature values from \cite{BenzAsphaug1999} (blue) and \cite{LeinhardtStewart2009} (orange) are shown as a reference. This illustrates that our weakest material ($Q_0$) is slightly weaker than the weak ice from \cite{LeinhardtStewart2009} while our strongest material ($100 Q_0$) is slightly stronger than the basaltic material from \cite{BenzAsphaug1999}.}
      \label{fig:Qstar}
\end{figure}

With regards to $Q^*_D$ we currently do not have a clear sense of the collisional strength of Trojans but can turn to literature for guidance.
How impacts disrupt a body depends on the size of the target body, the mass ratio of the target and the impactor, the relative speed and angle of impact, as well as the material properties of the bodies \citep[e.g][]{Holsapple1986,BenzAsphaug1999,michel2001,LeinhardtStewart2009,Jutzi2010,Benavidez2012,Holsapple2019,Jutzi2019}.

For the functional form of all $Q^*_D$ we follow the one in \cite{BenzAsphaug1999} where
\begin{equation}\label{eq:Qstar}
    Q^*_D = C_s\left(\frac{D}{2}\right)^{s_s} + \rho C_g \left(\frac{D}{2}\right)^{s_g} \quad,
\end{equation}
with $D$ being the diameter of the parent body in cm, $\rho$ the bulk density of the body, and $C_{s,g}$ and $s_{s,g}$ are the scaling constants and slopes in the strength and gravity regimes.
Different disruption laws are shown in the left panel of Figure~\ref{fig:Qstar}.
They have primarily been derived for properties suitable to rocky or icy objects.
Different techniques have been used to determine $Q^*_D$.
First, numerical models, as e.g. Smoothed Particle Hydrodynamics, SPH \citep[e.g.][]{BenzAsphaug1999,michel2001,Jutzi2010}, have been used to simulate the collision of bodies at different impact speeds and angles.
Second, models for the collisional evolution of populations (akin to \textit{Boulder}) have been used to constrain the disruption laws by simulating the evolution of the SFD and comparing them with the currently observed SFD in an inverse problem sense \citep[e.g][]{Bottke2005,Bottke2020,Benavidez2022MNRAS}. 
This method, though it can provide insights into the disruption law of a population, cannot infer material properties on its own.
Third, ground truths from laboratory experiments \citep[e.g.][]{SenftStewart2007,SenftStewart2008} are crucial to determine the disruption or cratering laws. 

The $Q^*_D$ illustrated in Figure~\ref{fig:Qstar} show many common features.
They all have a minimum at a similar size of $D_{min}\in[200,400]$~m.
This divides the curves into the so-called gravity regime (to the right of the minimum) and the strength regime (to the left of the minimum).
The slopes in both the gravity and strength regime do not vary strongly.
Lastly, despite the similarity in shape, the $Q^*_D$ functions vary significantly in absolute strength.
The weak ice material of \cite{LeinhardtStewart2009} (orange line in Fig.~\ref{fig:Qstar}) is almost two orders of magnitude weaker than the basaltic material in \cite{BenzAsphaug1999} (blue line in Fig.~\ref{fig:Qstar}).

We have for this reason chosen to retain the canonical shape of $Q^*_D$ and vary only the magnitude over two orders of magnitude.
Our weakest material will be denoted by $Q_0$ (see right panel of Fig.~\ref{fig:Qstar}) and corresponds to materials slightly weaker than the weak ice from \cite{LeinhardtStewart2009}.
We then include scaling of $2.2 Q_0$, $4.6 Q_0$, $10 Q_0$, $22 Q_0$, $46 Q_0$, and $100 Q_0$ where the latter is slightly stronger than the basaltic material from \cite{BenzAsphaug1999}.
Given the match in the familiy SFD with the runs of \cite{Benavidez2012} (Fig.~\ref{fig:EurybatesSFD}), the strength of Proto-Eurybates would correspond to our $4.6 Q_0$ case.

Finally, we will also consider a more exotic $Q^*_D$ (purple line in right panel of Fig.~\ref{fig:Qstar}) which follows the basaltic material from \cite{BenzAsphaug1999} in the gravity regime but then continues down to $D_{min}=20$~m before transitioning to the strength regime.
We will discuss the motivation and implications for this kind of $Q^*_D$ in Sec.~\ref{sec:results}.

\subsection{Input parameters}
In the previous section, we have mentioned the input parameters that are needed for this work.
First, we have described that the \textit{Boulder} code requires population-specific parameters.
These are the initial SFD, intrinsic impact probability, $P_i$, and the average impact speeds, $v_i$, of the population.
In the case of the JTs we can treat the two swarm separately because they do not overlap dynamically and thus do not contribute to the collisional evolution of each other.
For the SFD we, therefore, assume a slightly larger population than L$_4$.
The largest object in our SFD has a diameter of 140~km.
Between our largest size and 100~km we have a steep slope leading to 15 Trojans larger than 100~km.
Below 100~km we impose a cumulative slope of 2.1 down to our smallest size of 2~m.
This initial SFD is shown in later figures (e.g. Fig.~\ref{fig:EvolvedTrojanSFD}) and has a total mass of $4\times10^{-6}$~M$_{\bigoplus}$.
This mass is consistent with the estimate by \cite{Nesvorny2013ApJ}.
The collisional environment within each swarm is defined by $P_i = 7\times10^{-18}$~km$^{-2}$~yr$^{-1}$ and $v_i = 4.6$~km~s$^{-1}$ \citep[][]{Davis2002,Nesvorny2018NatAs}.
The intrinsic impact probability takes into account the resonant dynamics of the Trojans.

The second set of input parameters relates to the material properties of the JTs through the $Q^*_D$.
As described above we have tested two types of disruption laws, one with a canonical shape and a minimum in $Q^*_D$ at $D=250$~m and one with a minimum at $D=20$~m.
The former $Q^*_D$ form a set with the weakest material labeled $Q_0$. 
Scaled versions of $Q_0$ have been tested up to $100\times Q_0$ (see Fig.~\ref{fig:Qstar}).
The parameters $Q^*_D$ (Eq.~\ref{eq:Qstar}) used in this work are listed in Table~\ref{tab:QstarParmeter}.

\begin{table}[]
\caption{Parameters used for $Q^*_D$ in this work, and resulting minimum of $Q^*_D$, $D_{min}$.}
\label{tab:QstarParmeter}
\begin{tabular}{rrrrrrl}\hline
 $C_s$              & $s_s$   & $C_g$                     & $s_g$  & $\rho$         & $D_{min}$ & description \\
 $[$erg g$^{-1}]$   &         & $[$erg cm$^{3}$ g$^{-2}]$ &        & $[$g cm$^{-3}]$& $[$m$]$   &\\ \hline\hline
 $1.50\times 10^6$   & $-0.40$ & $0.05$                    & $1.30$ & $1$            & $250$     & $Q_0$ case \\
 $7.65\times 10^5$   & $-0.36$ & $1.40$                    & $1.36$ & $1$            & $20$      & $Q^*_D$ with minimum at $20$~m \\ \hline
\end{tabular}
\end{table}

\section{Results}\label{sec:results}
There are essentially three unknowns in our simulations as defined above: 1) the initial Jupiter Trojan SFD below 10~km; 2) the collisional strength of JTs, $Q^*_D$; and 3) the age of the Eurybates family.
Here we present our results that will illustrate how these three properties affect each other, and what we can learn about them.
With respect to the age of the Eurybates family, we stipulate that this is equal to the age of Queta, thereby assuming that Queta formed during the family forming collision.
While this is not necessarily true we are of the opinion that this is a sound assumption absent any other data.
As a result we use the age of the family and Queta interchangeably. 

To keep things simple, we first start with a simplified situation of a static Jupiter Trojan SFD and will then add more complexity by allowing the SFD to collisionally grind down over time.
Finally, we present the results for a non-canonical $Q^*_D$ which might have interesting implications for other small body populations in the outer Solar System.

\subsection{Static SFD}\label{sec:results-staticSFD}

In a first step we assume that the SFD of the JT population has not evolved over the age of the Solar System.
To do this we switch off the collisional evolution of the SFD in \textit{Boulder}.
This retains only the part of the code that monitors impacts on the Eurybates-Queta system and as a consequence allows us to separate the effects from the evolving Trojan SFD (see next section) and the inherent survival probability of Queta given a certain Trojan population.
Though this is a simplification it does allow us to retrieve a first estimate for the time scales.
As described above we assume two different initial SFDs.

\begin{figure}[ht]
      \centering
      \includegraphics[width=\textwidth]{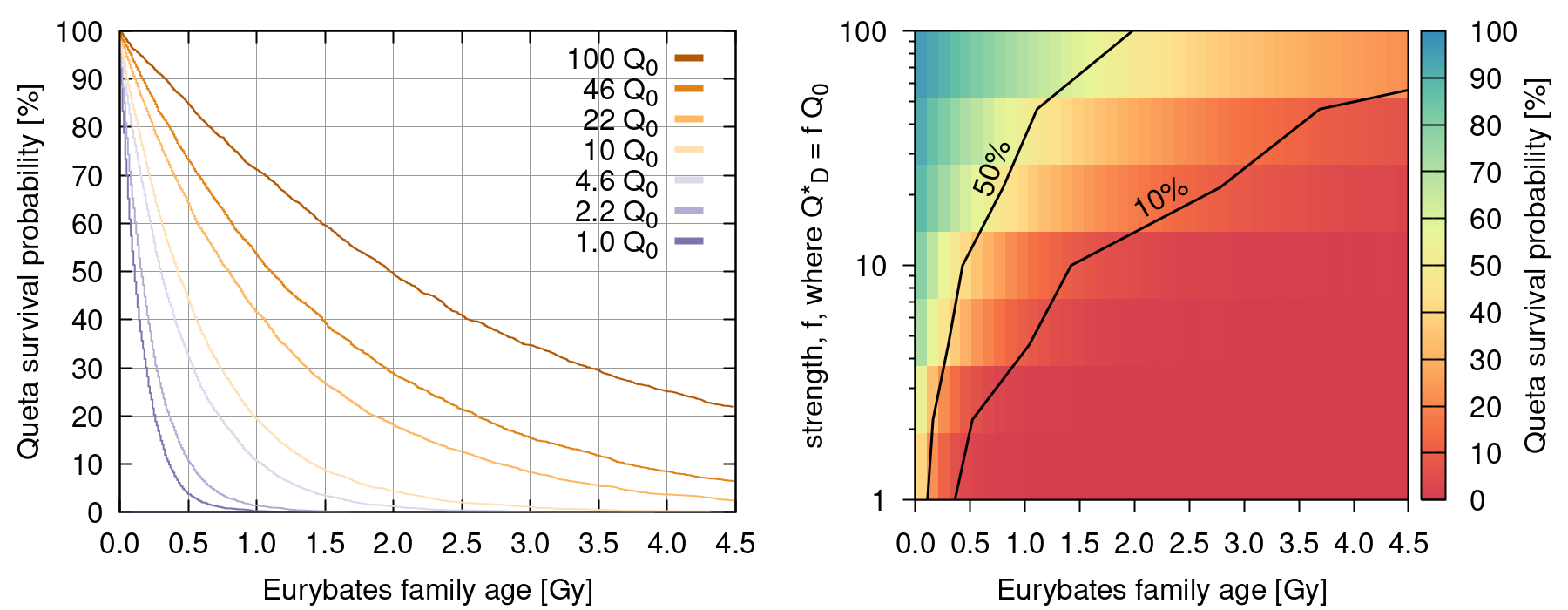}
      \caption{The left panel shows the survival probability of Queta as a function of the family age for the different $Q^*_D$ of our study. The right panel shows the same information but as a function of collisional strength. The two black lines show the $10\%$ and $50\%$ survival contours. Cases below the $10\%$ contour should be discarded as viable cases. For these cases a static SFD with a cumulative slope of $2.1$ between 2~m and 100~km was assumed.}
      \label{fig:QuetaDecay}
\end{figure}

First, we assume that the SFD below 10~km continues with the same slope as at larger sizes (2.1, see orange line in Fig.~\ref{fig:EvolvedTrojanSFD}).
In this case the probability that Queta survives is a simple exponential decay over time and depends only on $Q^*_D$ (left panel of Fig.~\ref{fig:QuetaDecay}).
At a certain age of the family the probability that Queta would have survived to this day becomes so small that we should not consider these cases.
For our purpose we use a $10\%$ survival probability as a discriminator between a scenario which we still consider plausible and scenarios where it would simply be too unlikely for Queta to have survived.
We find that a large part of parameter space can be excluded (right panel of Fig.~\ref{fig:QuetaDecay}).

Therefore, Figure~\ref{fig:QuetaDecay} can answer the question of the age of the Eurybates family as a function of collisional strength.
Any cases below the $10\%$ contour line in the right panel of Fig.~\ref{fig:QuetaDecay} should be considered statistically improbable.
Given this static SFD with a slope of $2.1$ between 2~m and 100~km, we find that $Q^*_D$ needs to be strictly larger than $4.6 Q_0$.
Otherwise Queta would not survive for more than 1~Gy, the lower limit for the family age.
Interestingly, $4.6 Q_0$ corresponds to the $Q^*_D$ found by \cite{Benavidez2018} for their rubble pile asteroids.
As we have seen above, one of their simulations is a good fit to the SFD of the Eurybates family.
If the parent body of the Eurybates family was indeed this weak ($4.6 Q_0$), and this disruption law is representative of Jupiter Trojans, and the current day SFD continues with the same slope down to small sizes, then this result would suggest that the Eurybates family is a ``young'' family (1 Gy).

We also observe that the age of the family and the collisional strength are directly linked.
As the collisional strength increases so does the age of the family.
Strictly speaking these are all upper limits for the age of the family.
For example, our strongest material ($100 Q_0$) never falls below $10\%$ and hence the family age could be up to the age of the solar system.
On the other hand the $46 Q_0$ case, which in fact corresponds roughly to the $Q^*_D$ of basaltic material in \cite{BenzAsphaug1999} (see blue line in Fig.~\ref{fig:Qstar}) gives us an upper limit of $\sim3.7$~Gy for the family age.

\begin{figure}[ht]
      \centering
      \includegraphics[width=\textwidth]{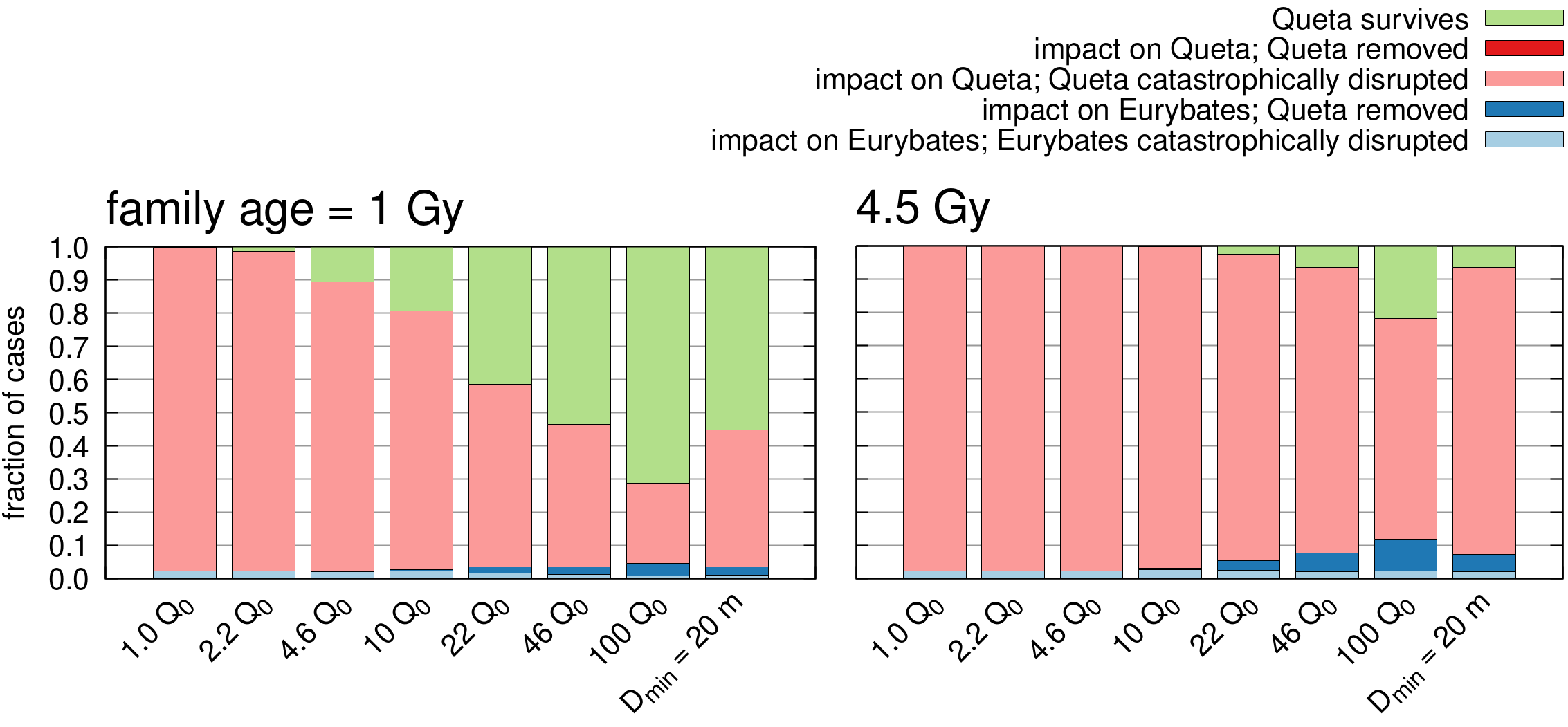}
      \caption{The fraction of outcomes for the Eurybates-Queta system are shown for a family age of 1 Gy (left panel), and 4.5 Gy (right panel). A static SFD with a slope of $2.1$ between 2~m and 100~km (section~\ref{sec:results-staticSFD}) is assumed here.). A total of 1000 runs were performend to estimate the fraction of outcomes of each case.}
      \label{fig:QuetaDiscardTypesStat}
\end{figure}

We can also examine what contributes to the destruction of the Eurybates-Queta system.
Fig.~\ref{fig:QuetaDiscardTypesStat} shows our results for different $Q^*_D$ and a family age of 1~Gy (left) and 4.5~Gy (right).
We find that in the overwhelming fraction of cases, an impact occurs on Queta that catastrophically disrupts it (pink in Fig.~\ref{fig:QuetaDiscardTypesStat}).
There are no cases where an impact on Queta perturbs its orbit enough for the binary to be dissolved (red in Fig.~\ref{fig:QuetaDiscardTypesStat}).
In a minor fraction of cases (up to $10\%$ after 4.5~Gy for $100 Q_0$), an impact on Eurybates leads to the ejection of Queta.
It is interesting to note that the relative fraction of cases where an impact on Eurybates destroys Eurybates (light blue in Fig.~\ref{fig:QuetaDiscardTypesStat}) decreases with respect to when such an impact dissolves the binary (dark blue in Fig.~\ref{fig:QuetaDiscardTypesStat}).
This result makes sense because as the material strength increases it becomes harder to catastrophically disrupt Eurybates but ``easier'' to nudge Eurybates such that the binary orbit is dissolved and Queta lost from the system.
The fact that the survival of Euybates-Queta system is dominated by impacts on Queta is noteworthy because it is different than what has been observed for other binaries, in particular the large Trojan binary pair of Patroclus and Menoetius \citep[P-M binary;][]{Nesvorny2018NatAs}, the final fly-by target of the Lucy mission.
The P-M system is an almost equal size binary with diameters of 113 km and 104 km respectively.
The survival of the P-M binary is limited by impacts on one of the components that alter the orbit such that the binary is dissolved.
The binaries survival is thus sensitive to the semi-major axis of the binary \citep[][]{Nesvorny2018NatAs} because binaries with larger semi-major axis are dissolved more easily than tight binaries.
\cite{Nesvorny2018NatAs} estimated that the specific energy needed to dissolve the P-M binary through a non-catastrophic impact is much smaller than the respective $Q^*_D$.
In our case, it is much easier to catastrophically disrupt the much smaller Queta ($\sim1$~km).
Therefore, in contrast to the  P-M binary, the Eurybates-Queta system's survival is not sensitive to the semi-major axis.

\begin{figure}[ht]
      \centering
      \includegraphics[width=0.5\textwidth]{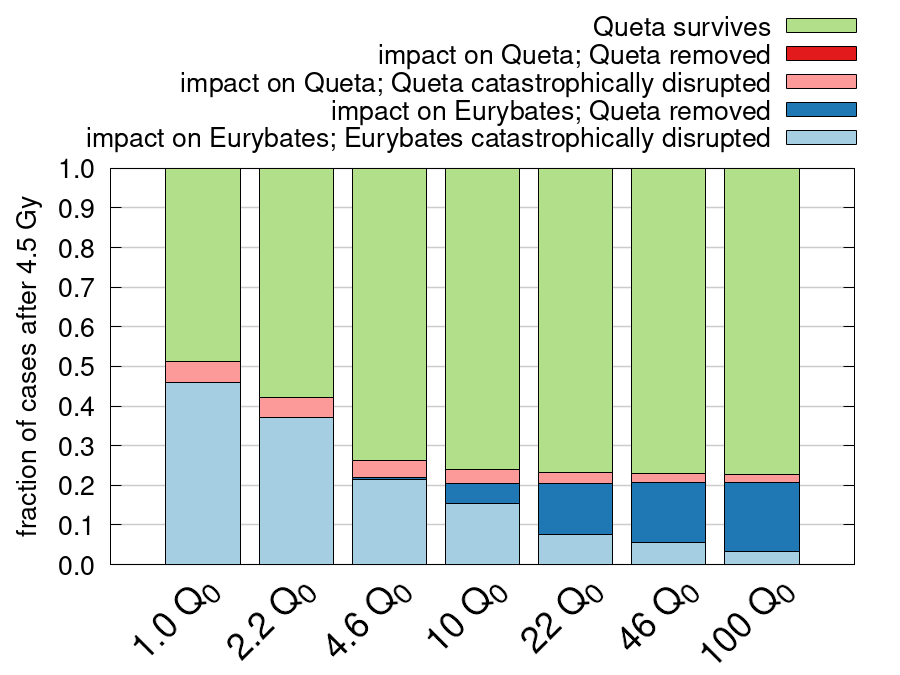}
      \caption{The fraction of outcomes for the Eurybates-Queta system are shown for a family age of 4.5 Gy when the initial Jupiter Trojan SFD is shallow (power law of 1) below 10~km. }
      \label{fig:QuetaDiscardTypes-2.0}
\end{figure}

Up to now we have assumed that the Jupiter Trojan SFD simply continues below 10~km with respect to the slope above 10~km.
Because there are indications that the SFD bends/breaks at or somewhat below 1~km \citep[][]{YoshidaNakamura2008,WongBrown2015,YoshidaTerai2017}, we now assume that the Trojan SFD has a break at 10~km and continues at small sizes with a slope of $q=1$ (significantly flatter than above 10~km).
In this case, we find a very different picture to the one above (Fig.~\ref{fig:QuetaDiscardTypes-2.0}).
We find that the age and collisional strength become unconstrained.
Queta can survive 4.5~Gy in all cases of $Q^*_D$.
In the worst case scenario with extremely weak material ($Q_0$) there is still a 50/50 chance of Queta surviving.
In most cases ($Q^*_D \ge 4.6Q_0$) the survival probability of Queta over 4.5~Gy is $\sim80\%$.

This behavior is readily explained.
Due to the very flat SFD below 1~km, the impactor population that could destroy Queta is simply not very large.
Consequently the probability for Queta to be catastrophically disrupted (pink in Fig.~\ref{fig:QuetaDiscardTypesStat} and \ref{fig:QuetaDiscardTypes-2.0}) shrinks to the single percent digits.
In this situation, the probability of the Eurybates-Queta system being disrupted is dominated by impacts on Eurybates.
This happens either by direct disruption of Eurybates in the case of weak material or dissolution of the binary for strong material.
Further, we observe a transition from disruptions of Eurybates dominating the survival probability at low collisional strength to impacts on Eurybates dissolving the binary at large collisional strengths.
At low strength, Eurybates is easily disrupted and thus becomes the primary mode limiting the retention of Eurybates-Queta system.
As the strength increases Eurybates becomes hard to disrupt but can be nudged strongly enough to dissolve the orbit of Queta.

\subsection{Evolving SFD}\label{sec:results-evolvingSFD}
\begin{figure}[ht]
      \centering
      \includegraphics[width=\textwidth]{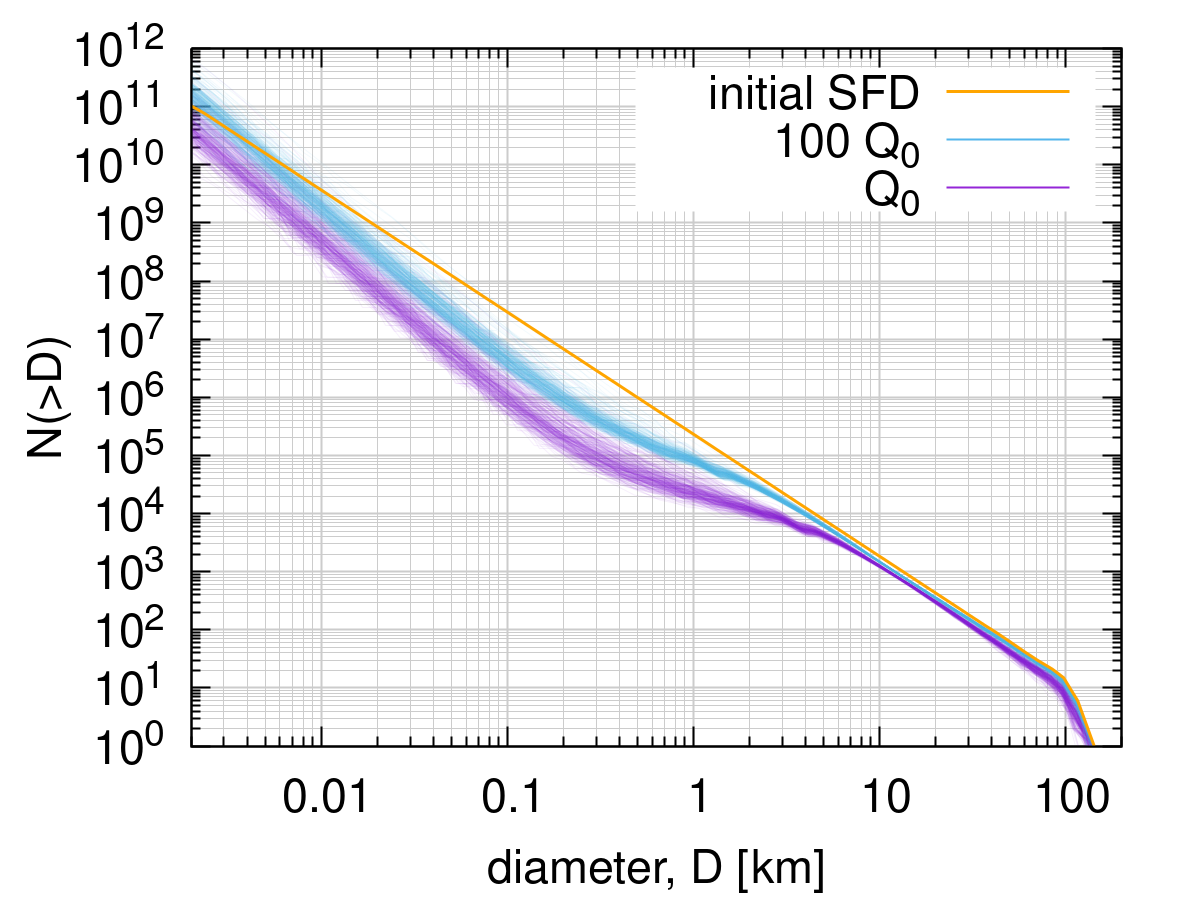}
      \caption{The initial Trojan SFD is shown in orange. The faint lines in purple are 300 different SFDs after 4.5~Gy of collisional grinding assuming $Q^*_D=Q_0$ while the blue lines assume $Q^*_D=100Q_0$.}
      \label{fig:EvolvedTrojanSFD}
\end{figure}

As argued above the assumption of a static Jupiter Trojan SFD over the age of the Solar System is likely to be an unrealistic simplification.
Accordingly, we now consider an initial Trojan SFD with a slope of 2.1 below 10~km (see orange line in Fig.~\ref{fig:EvolvedTrojanSFD}) and let it evolve over the age of the Solar System.
The results of 300 such simulations for a $Q^*_D$ of $Q_0$ and $100Q_0$ are shown in Figure~\ref{fig:EvolvedTrojanSFD}.
We show 300 simulations to illustrate the stochastic nature of the collisional evolution of the Trojan population.
The sets of simulations differ only in the random seed given to the code.
Each simulation, therefore, signifies a possible future of the same initial state of the system.

\begin{figure}[ht]
      \centering
      \includegraphics[width=\textwidth]{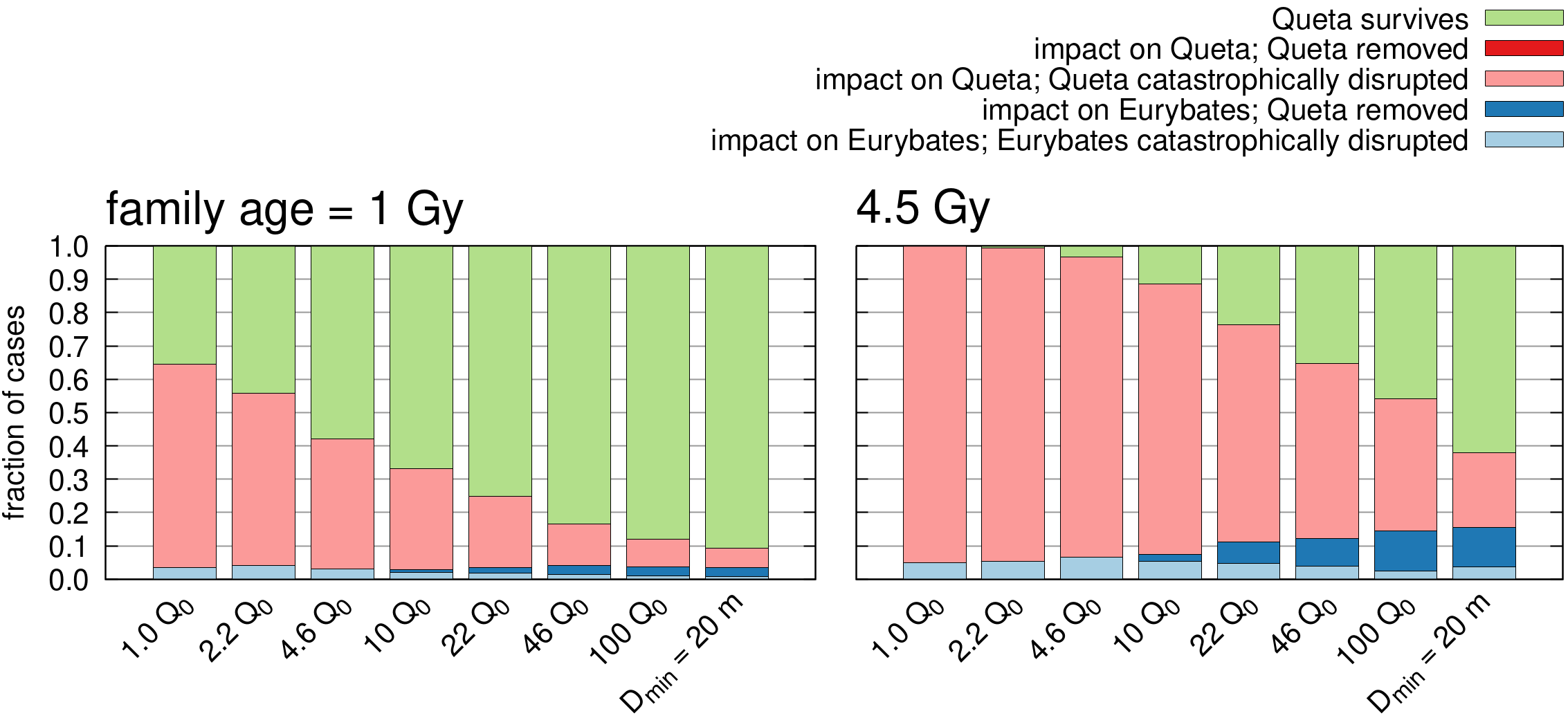}
      \caption{The fraction of outcomes for the Eurybates-Queta system are shown for a family age of 1 Gy (left panel), and 4.5 Gy (right panel). The results here take into account the changing SFD due to collisional grinding (section~\ref{sec:results-evolvingSFD}). A total of 1000 runs were performend to estimate the fraction of outcomes of each case. FIGURE WAS SPLIT FROM FIG.~\ref{fig:QuetaDiscardTypesStat}.}
      \label{fig:QuetaDiscardTypesDyn}
\end{figure}

We find that the slope between $\sim 6$~km and 100~km remains unaltered though the population is depleted slightly in both cases.
In both cases shown in Fig.~\ref{fig:EvolvedTrojanSFD} the SFD below $\sim 6$~km is significantly affected by collisional grinding.
In particular, the population is depleted between 10~m and 3~km.
This results in a turnover between 2 and 6~km.
The location of this bend depends on the $Q^*_D$.
It is likely that this is the cause of the bend in the SFD seen by \cite{YoshidaNakamura2008} and \cite{WongBrown2015}. 
But we should also point out that \cite{YoshidaTerai2017} reported no detection of this bend and thus its existence and location requires further observations.
From a theoretical standpoint we expect there to be a bend in the Trojan SFD (Fig.~\ref{fig:EvolvedTrojanSFD}) therefore it is likely ``merely'' a matter of getting more observations to identify it.

The depletion of the JT population between 10~m and 1~km directly affects the survival probability of Queta because this is the impactor population responsible for its destruction (see above).
We find that the survival probability of Queta significantly increases in all cases (see bottom row of Fig.~\ref{fig:QuetaDiscardTypesDyn}).
Compared to the static SFD (Fig.~\ref{fig:QuetaDiscardTypesStat}), all of the $Q^*_D$ allow Queta to plausibly survive with a family age of more than 1~Gy.
But for these cases, we also need a material strength of at least $10 Q_0$ for Queta to have a 10\% chance of surviving the age of the Solar System.

\begin{figure}[ht]
      \centering
      \includegraphics[width=\textwidth]{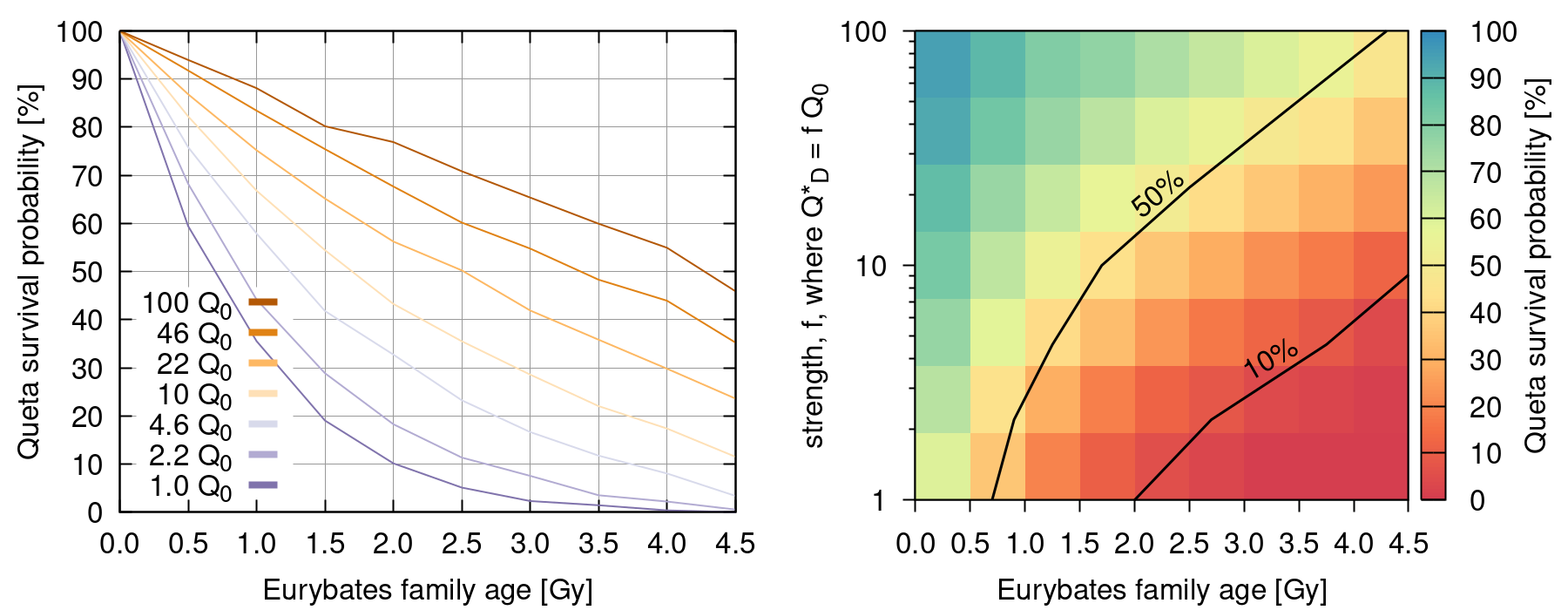}
      \caption{The left panel shows the survival probability of Queta {for the evolving Trojan SFD case} as a function of the family age for the different $Q^*_D$ of our study. The right panel shows the same information but as a function of collisional strength. The two black lines show the $10\%$ and $50\%$ survival contours. Cases below the $10\%$ contour should be discarded as viable cases.}
      \label{fig:QuetaDecayEvol}
\end{figure}

Figure~\ref{fig:QuetaDecayEvol} shows the survival probability as a function of the family age.
Compared to the static SFD (Fig.~\ref{fig:QuetaDecay}) we observe that the decay is no longer exponential but becomes almost linear in time for the strongest material.
Further, the part of parameter space consistent with Queta's existence (cases above the $10\%$ contour) is greatly expanded compared to the static SFD (right panel of Fig.~\ref{fig:QuetaDecay}).
This result is not a surprise because there are fewer projectiles in the population capable of destroying Queta.
The collisional strength derived for the \cite{Benavidez2012} rubble piles with $Q^*_D=4.6 Q_0$ now puts an upper limit on the family age.
The family is no older than $3.7$~Gy with a probability of $90\%$ for this size distribution and disruption law.

\subsection{Unconventional $Q^*_D$}\label{sec:results-unconventionalQstar}

Finally, we will examine what happens to our results if we employ a more unconventional $Q^*_D$.
As described in section~\ref{sec:results-evolvingSFD} (see also Fig.~\ref{fig:Qstar}) the minimum value of $Q^*_D$ in literature for rocky and icy targets is typically around a diameter of $D_{min}=200$~m.
There are indications from cratering results obtained by the New Horizons mission to the Pluto system and to the cold classical KBO Arrokoth \citep{Singer2019,Morbidelli2021}, however, that this might not hold for icy bodies of the outer Solar System. 
The critical clue is that the cumulative power law slope of the Kuiper belt SFD is approximately $q=1$ over the size range between 1 km and tens of meters in diameter \cite{Morbidelli2021}.  

As discussed in section~\ref{sec:results-evolvingSFD}, the location where the intermediate slope steepens back up to the slope in the strength regime is indicative of the minimum in $Q^*_D$.
For the asteroid belt, this change in slope is observed at 200 m \citep[e.g.][]{Bottke2020}. 
For the Kuiper belt, we have yet to identify the size of where this putative change in slope would take place. 
All we can say is that if it exists, it has to occur at sizes smaller than tens of meters.  Accordingly, we infer that $Q^*_D$ for Kuiper belt objects, and presumably Trojans, need to have a minimum value smaller than a few tens of meters.
We therefore assume an unconventional $Q^*_D$ (purple line in right panel of Fig.~\ref{fig:Qstar}) which follows the \cite{BenzAsphaug1999} basaltic material in the gravity regime but then continues down to include continuously weaker material strength all the way down to $D_{min}=20$~m before transitioning to the strength regime.
This minimum value is consistent with the arguments made by \cite{Morbidelli2021}, namely that a upturn near that point would provide the means to explain the quantity of dust observed in the Edgeworth-Kuiper belt by New Horizons.

\begin{figure}[ht]
      \centering
      \includegraphics[width=\textwidth]{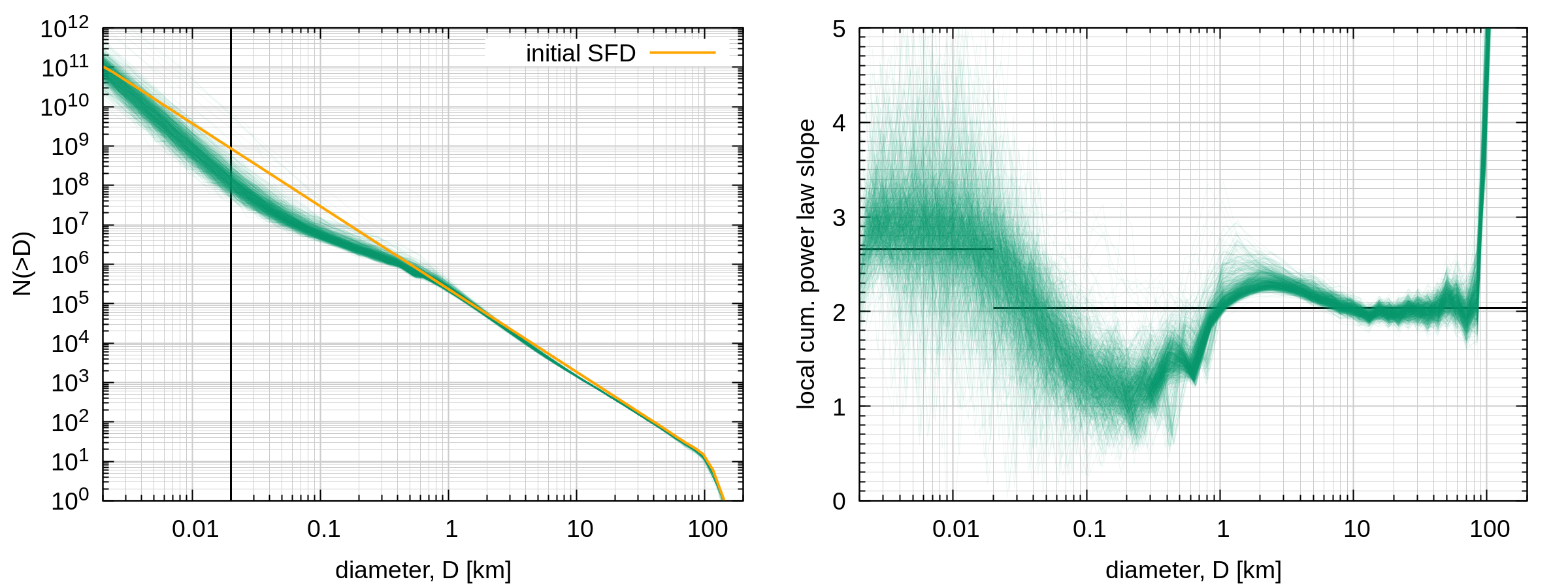}
      \caption{The left panel shows the evolved SFDs after 4.5~Gy assuming a $Q^*_D$ with $D_{min}=20$~m. For reference the initial SFD is shown in orange. The right panel shows the slopes assuming a local power law as well as the theoretical slopes (solid black) given by Eq.~\ref{eq:slopes}. Both the largest and smallest Trojans approach the predicted slopes. The intermediate size regime shows a mild wave of the SFD which steepens up between 1-10~km before becoming shallow below 1~km and finally converging to the strength regime slope.}
      \label{fig:MinQSFD}
\end{figure}

Figure~\ref{fig:MinQSFD} shows what the SFD evolved to after 4.5~Gy.
The SFD is more complex than the ones previously shown (Fig.~\ref{fig:EvolvedTrojanSFD}).
We interpret it using the local slopes shown in the right panel.
Here we have defined the local slope as the power law index connecting the two adjacent size bins within the Boulder output results. 
The typical distance between size bins is $1.25D$.

\cite{OBrienGreenberg2003} showed for a population in collisional equilibrium that the differential power law exponent, $p$, is connected to the slope, $s$, of the gravity or strength regime respectively via:
\begin{equation}\label{eq:slopes}
    p = \frac{7+\frac{s}{3}}{2+\frac{s}{3}} \quad.
\end{equation}
If $Q^*_D$ is independent of size (i.e. $s=0$) we retrieve the classical Dohnanyi steady-state solution of $p= 3.5$ \citep{Dohnanyi1969}.
In our case the gravity regime of $Q^*_D$ has a slope of $s=1.3$ resulting in $p=3.05$ which corresponds to a cumulative slope of $2.05$.
This is close to the observed slope between 10 and 100 km which could indicate that this part of the population was in collisional equilibrium when the Trojans were captured.
The slope at small sizes should corresponds to the respective slope predicted for the strength regime.
In between the two steep slopes at the very small and large sizes we find a shallow slope in the diameter range from roughly $30$~m to $1$~km.
Note that the location where the SFD transitions to the slope in the strength regime is precisely the location of the minimum in $Q^*_D$ at roughly $D_{min}=100-200$~m of the traditionally shaped $Q^*_D$ (Fig.~\ref{fig:EvolvedTrojanSFD}), and $D_{min}=20$~m for our unconventional $Q^*_D$ (Fig.~\ref{fig:MinQSFD}).
This makes sense as objects smaller than $D_{min}$ are stronger than their larger companions. 
As a consequence they can more easily break larger objects but not be easily broken up themselves.

Starting at large JTs ($10$~km~$<$~D~$<100$~km) in Fig.~\ref{fig:MinQSFD}, the SFD is largely unchanged from the initial SFD, and retains $q\sim2$. 
Few of these objects have disrupted.
At 10~km the slope begins to steepen slightly reaching a peak at around $2.5$~km before beginning to flatten.
This indicates a steady state “bump” is being formed by collisional evolution; it is analogous to the bump seen in the asteroid belt near 2-3 km, and we will discuss its origin below.
The flattest part of the SFD occurs at $\sim200$~m.
From there the slope steepens up again to reach the predicted slope in the strength regime for a collisionally-evolved Dohnyani-like SFD \citep[][]{OBrienGreenberg2003}.
This final slope is reached as expected around $D_{min}=20$~m.
As with the asteroid belt, the limited number of projectiles near this minimum means there should be an excess of bodies at larger sizes corresponding to the targets that would be disrupted by 20~m projectiles. 
That explains the change of the aforementioned ``bump'' at around 1~km. 
Similarly, the slight ``dip'' in the population around 10~km is produced by the excess number of 1~km bodies in the bump.
Intriguingly, there may be some limited evidence of a ``dip'' near 10 km Trojans.
The tentative shallowing of the slope at small sizes \citep[$\sim5$~km][]{YoshidaNakamura2008,WongBrown2015} is associated with a slightly steeper slope than the nominal $2.1$ just before the apparent turn over.
While this is not definitive proof, and details need to be worked out in future work and observations, we argue that there is sufficient support here to warrant seriously entertaining the possibility of an unorthodox $Q^*_D$ with a $D_{min}$ at $\sim20$~m.

It is also worth mentioning that the bump at $\sim1$~km can cause the current day population to exceed the initial population at that size (left panel of Fig.~\ref{fig:MinQSFD}). 
This is not observed in any of the cases that use a traditionally-shaped $Q^*_D$.
The reason is that traditionally shaped $Q^*_D$ produce a turnover at larger sizes (3-7 km) than for our unconventional $Q^*_D$ (below 1~km).
This leads to a smaller 1~km impactor population for traditional $Q^*_D$ functions that is not large enough to disrupt many Trojans of the order of 10~km, i.e., it cannot produce a dip at those sizes.
These larger breakups act as a source population for further 1~km Trojans and, therefore, allow a ``bump'' in the unconventional $Q^*_D$ case.

This model not only has the advantage of fitting into the data returned by the New Horizons mission \citep{Singer2019,Morbidelli2021} and qualitatively matching the flattening of the Jupiter Trojan SFD at small sizes \citep{YoshidaNakamura2008,WongBrown2015} but it also liberates the constraint on the collisional lifetime of Queta and thus the age of the Eurybates family.

In fact, this unconventional $Q^*_D$ would allow Queta to survive under any of our assumptions for at least the age of the Solar System (Figs.~\ref{fig:QuetaDiscardTypesStat} and \ref{fig:QuetaDiscardTypesDyn}).
The reasons is that the projectile population capable of disrupting Queta is low and stays that way for 4.5 Gyr.

This unconventional $Q^*_D$, though, would not match the strength of Proto-Euryabes derived from SPH simulations in Sec.~\ref{sec:EurybatesFamily}.
Note also that we have tested smaller values for $D_{min}$, but they produce SFDs that are less consistent with New Horizons crater constraints because the shallow slope in the SFD continues to even smaller sizes than shown in Fig.~\ref{fig:MinQSFD}. 
This would make it difficult to reconcile the New Horizons crater record with the Kuiper belt dust observations \citep[see argument in][]{Morbidelli2021}.

\section{Discussion}\label{sec:discussion}
The main avenue for disrupting the Euybates-Queta system is via direct impacts on Queta (Figs.~\ref{fig:QuetaDiscardTypesStat} and \ref{fig:QuetaDiscardTypesDyn}).
Thus the crucial factor in the survival of Queta is the impactor population below 1~km. 
For Queta to survive from its creation in the Eurybates family forming impact to today and not have its survival be a fluke, we need to limit the number of projectiles that can disrupt Queta.  
From a collisional evolution perspective, that means the slope of the Jupiter Trojan SFD below 1~km needs to be shallow.
This can be achieved by material with a traditionally shaped $Q^*_D$ or a more unorthodox shaped $Q^*_D$ that has a minimum at 20~m.

In the first case, not all $Q^*_D$ functions allow Queta to survive the age of the Solar System.
If the collisional strength is too low, Queta is too easily disrupted, even though the Trojan SFD collisionally grinds down a substantial amount.
For these kinds of runs, a minimum of $10Q_0$ is needed for the long term (i.e. the age of the Solar System) survival of Queta.
The problem is that this value is inconsistent with the inferred strength against disruption we have found for Eurybates based on the family SFD ($4.6Q_0$; Fig.~\ref{fig:EurybatesSFD}) derived from the SPH family forming impact simulation.
If, on the other hand, we relax the condition that the family be as old as 4.5~Gy and instead insist that $Q^*_D=4.6Q_0$ then there is a $90\%$ chance that the family is no older than 3.7~Gy (Fig.~\ref{fig:QuetaDecayEvol}).
Thus for a traditional $Q^*_D$, either, the familiy is likely younger than the age of the Solar System, or the collisional strength of Proto-Eurybates ($4.6Q_0$) was lower than the overall Trojan population ($>10Q_0$).

The case of a traditionally shaped $Q^*_D \le 4.6 Q_0$ leads to a high probability that Queta experiences a catastrophic impact during the lifetime of the Solar System.
For Queta to still survive in such a scenario would require an additional mechanism.
It is imaginable that Queta re-accumulated after a catastrophic disruption event, or that Queta was initially much larger than today and has undergone significant collisional grinding since.
The latter, in particular, seems unlikely as impact simulations primarily result in small satellites \citep[e.g.][]{Durda2004}. 
The existence of small satellites also seems consistent with observed satellites in the main belt \citep[e.g. Fig.~4 in ][]{Noll2020}.

In the second case, with an unconventional shaped $Q^*_D$ that has its minimum at 20~m, Queta can easily survive the age of the Solar System and, therefore, the Eurybates family can be as old as 4.5~Gy.
But, again, this would bring the collisional strength of Proto-Eurybates and the overall Trojan population into disagreement.

The fact that several $Q^*_D$ are consistent with the survival of Queta, raises the question as to how we shall distinguish these cases.
Figures~\ref{fig:EvolvedTrojanSFD} and \ref{fig:MinQSFD} show that the Jupiter Trojan SFD can evolve to very different shapes when a variety of $Q^*_D$ functions are used.
This result has direct consequences for the expected crater SFD that the Lucy mission will observe.
A back of the envelope estimate shows that these contrasting Trojan SFDs lead to a stark differences in the expected number of craters forming on a Trojan over a given amount of time, e.g., Eurybates (Table~\ref{tab:cratersEurybates}).
For the purpose of this estimate, we assume a crater to impactor ratio of 10 \citep[e.g.][]{Singer2019} (e.g. a 1~km impactor would produce a 10~km crater).
As Table~\ref{tab:cratersEurybates} illustrates the number of craters on Eurybates increases by a factor of four when the collisional strength increases from $Q_0$ to $100Q_0$.
The fact that our unconventional $Q^*_D$ results in the largest number of craters seems counterintuitive at first but Figs.~\ref{fig:EvolvedTrojanSFD} and \ref{fig:MinQSFD} hold the answer.
Because the shallow part of the Trojan SFD occurs at smaller sizes for the unconventional $Q^*_D$ compared to the traditionally shaped $Q^*_D$, the overall population at sizes between 100~m and 1~km, corresponding to a projectile size range between 10 m and 100 m, is larger. 
This outcome leads to more craters.
Additionally, the case of our unconventional $Q^*_D$ likely leads to a saturated cratered surface, potentially complicating how our crater SFDs will be linked to the impactor SFD.
Furthermore, the ratio between the number of craters larger than 1~km and 10~km is different for the case of our unconventional $Q^*_D$ compared to the traditionally shaped $Q^*_D$ with the latter producing more small craters for every large crater.
In the case of traditionally shaped $Q^*_D$ the slope of the Trojan SFD increases at roughly 300~m while for the unconventional $Q^*_D$ the slope remains flat between 60~m and 1~km.
This results in the different ratios of large to small impacts.
Lucy, being able to measure craters as small as 70~m, and therefore sensitive to projectile sizes as small as 7~m, will be able to determine the crater SFD to sufficient precision to potentially differentiate between these case studies, assuming the surfaces are not saturated. 
In turn, this would allow us to indirectly estimate the collisional strength of the Trojans and the $Q^*_D$ disruption law followed by both Trojans and similar objects.

\begin{table}[]
\caption{Minimum number of expected craters on Eurybates, $N_C$, accumulated during 2~Gy which are larger than a certain diameter for different $Q^*_D$. This calculation does not take into account saturation.}
\label{tab:cratersEurybates}
\begin{tabular}{r|rrr}
$Q^*_D$ & $N_C(>1$~km$)$ &  $N_C(>10$~km$)$&  $\frac{N_C(>1~km)}{N_C(>10~km)}$\\\hline
$Q_0$           &  $200$ & $4$  & $50$\\
$100Q_0$        &  $800$ & $16$ & $50$\\
$D_{min}=20$~m  & $1200$ & $50$ & $24$
\end{tabular}
\end{table}

\section{Summary \& Conclusions}\label{sec:summaryConclusions}
In this work we have modeled the collisional evolution of the Jupiter Trojans and determined under which conditions the Eurybates-Queta system survives.
We have shown that the collisional strength of the Jupiter Trojans and the age of the Eurybates family and Queta are correlated.
This correlation depends itself on the initial Trojan SFD and how it collisionally evolves over time.

We found that the Eurybates family SFD matches best the outcome of SPH models of a $100$~km rubble pile target \citep{Benavidez2012} being hit by a $\sim35$~km impactor ($\log(M_{tar}/M_{imp})=1.4$) at $6$~km/s, and an impact angle of $75^{\circ}$.
This corresponds to objects with a collisional strength of $Q^*_D=4.6 Q_0$, where $Q_0$ corresponds to slightly weaker material than the ``weak ice'' from \cite{LeinhardtStewart2009}.

In the case of static SFDs that don't evolve over time we examined a nominal case where the Jupiter Trojan SFD at small sizes ($<10$~km) has the same slope as between 10-100~km.
With this SFD, and assuming the parent body of the Eurybates family was indeed weak, and is representative of Jupiter Trojans, we found that the Eurybates family must be a ``young'' family (1 Gy).

Should the Trojan SFD have break at 1~km and have a shallow slope below that break then we cannot constrain the age of the Eurybates family and collisional strength of the Trojans.
The slope of the SFD below 1~km is the driving property to assess the family age and collisional strength.
This is due to the fact that the population between 10~m and 1~km is the impactor population which can disrupt Eurybates and Queta.

Further, we found that the collisional grinding of the JT population results in a SFD that remains largely unaltered at large sizes ($>10$~km) but is then depleted at intermediate small sizes (10~m to 1~km).
This implies a turn over in the SFD, the location of which depends on the $Q^*_D$.
It is to be expected that the Trojan SFD bends towards a shallower slope between 1 and 10~km \citep{YoshidaNakamura2008,WongBrown2015,YoshidaTerai2017}.

For more realistic cases where the Trojan population is allowed to collisionally evolve we find the following:
A material strength of at least $10 Q_0$ is needed for Queta to survive the age of the Solar System.
For the likely strength of Eurybates ($4.6 Q_0$) there is a $90\%$ chance that the family cannot be older than $3.7$~Gy.

We find that an unconventional $Q^*_D$ with a minimum at $D_{min}=20$~m is a plausible candidate that does not require any additional assumptions on the Trojan SFD or age of the Eurybates family.
Further, it fits into the data returned by the New Horizons mission \citep{Singer2019,Morbidelli2021} for the craters on Charon and KBO Arrokoth as well as the qualitative behavior of the Jupiter Trojan SFD at small sizes \citep{YoshidaNakamura2008,WongBrown2015}.

Finally, we have shown how different $Q^*_D$ will impact the expected number of craters on the targets of the Lucy mission.
The data from Lucy may be able to differentiate between different cases of $Q^*_D$ and subsequently indirectly determine the collisional strength of Jupiter Trojans.

\newpage

\section{Appendix}\label{sec:appendix}

\setlength{\LTcapwidth}{\textwidth}

\begin{table}[ht]
\caption{List of all 400 Eurybates family members identified by the hierarchical clustering method (HCM) including the likely interloper (5258) Rhoeo (see discussion in Sec.~\ref{sec:constraints}). The table gives the asteroid number ($\#$), the provisional designation (prov. des.), the absolute magnitude (H), the diameter (D), as well as the proper semi-major axis ($a_{prop}$), eccentricity ($e_{prop}$), and inclination ($i_{prop}$). The source number for each property is given in brackets. A machine readable version of this table is available on https://zenodo.org/SOME-LINK\\
$\left[1\right]$ Minor Planet Center; 2020-12-08, https://minorplanetcenter.net//iau/lists/JupiterTrojans.html\\
$\left[2\right]$ NEOWISE data v2.0, \cite{Mainzer2019}, https://sbn.psi.edu/pds/resource/neowisediam.html\\
$\left[3\right]$ Mira Bro\v{z} \citep{Holt2020}}
\label{tab:EurybatesFamily}
\begin{tabular}{rllcccc}

nr	&	prov. des.	&	H$^{[1]}$	&	D [km]$^{[2]}$	&	$a_{prop}$ [au]$^{[3]}$	&	$e_{prop}$$^{[3]}$	&	$i_{prop}$ [$^{\circ}$]$^{[3]}$\\
3548	&	1973 SO	&	9.85	&	63.885 $\pm$ 0.299	&	5.29758 $\pm$ 0.00065	&	0.04351 $\pm$ 0.00014	&	7.4150 $\pm$ 0.0012 \\
5258	&	1989 AU1	&	10.33	&	53.275 $\pm$ 4.429	&	5.28801 $\pm$ 0.00115	&	0.05898 $\pm$ 0.00016	&	7.0285 $\pm$ 0.0018 \\
8060	&	1973 SD1	&	10.95	&	37.873 $\pm$ 0.567	&	5.29164 $\pm$ 0.00054	&	0.05452 $\pm$ 0.00119	&	7.3082 $\pm$ 0.0006 \\
9818	&	6591 P-L	&	11.07	&	28.076 $\pm$ 3.215	&	5.28832 $\pm$ 0.00114	&	0.04650 $\pm$ 0.00020	&	7.4624 $\pm$ 0.0092 \\
13862	&	1999 XT160	&	11.6	&	24.835 $\pm$ 0.589	&	5.29245 $\pm$ 0.00012	&	0.04372 $\pm$ 0.00003	&	7.3313 $\pm$ 0.0004 \\
18060	&	1999 XJ156	&	11.12	&	36.431 $\pm$ 3.966	&	5.29238 $\pm$ 0.00007	&	0.04448 $\pm$ 0.00003	&	7.4139 $\pm$ 0.0004 \\
24380	&	2000 AA160	&	11.2	&	31.607 $\pm$ 0.266	&	5.29533 $\pm$ 0.00037	&	0.04358 $\pm$ 0.00006	&	7.3448 $\pm$ 0.0007 \\
24420	&	2000 BU22	&	11.45	&	21.723 $\pm$ 1.211	&	5.30505 $\pm$ 0.00082	&	0.04881 $\pm$ 0.00022	&	7.2478 $\pm$ 0.0045 \\
24426	&	2000 CR12	&	12.13	&	14.336 $\pm$ 1.007	&	5.30301 $\pm$ 0.00057	&	0.03817 $\pm$ 0.00041	&	7.3780 $\pm$ 0.0022 \\
28958	&	2001 CQ42	&	12.18	&	21.577 $\pm$ 0.652	&	5.29703 $\pm$ 0.00102	&	0.03779 $\pm$ 0.00016	&	7.4208 $\pm$ 0.0011 \\
39285	&	2001 BP75	&	12.49	&	17.602 $\pm$ 0.499	&	5.29535 $\pm$ 0.00054	&	0.04695 $\pm$ 0.00009	&	7.4102 $\pm$ 0.0008 \\
39795	&	1997 SF28	&	12.42	&	18.342 $\pm$ 0.742	&	5.29306 $\pm$ 0.00012	&	0.05088 $\pm$ 0.00006	&	7.3939 $\pm$ 0.0003 \\
43212	&	2000 AL113	&	12.2	&	19.212 $\pm$ 1.09	&	5.28911 $\pm$ 0.00082	&	0.04813 $\pm$ 0.00021	&	7.2695 $\pm$ 0.0059 \\
43436	&	2000 YD42	&	12.12	&		&	5.29723 $\pm$ 0.00032	&	0.05201 $\pm$ 0.00007	&	7.1651 $\pm$ 0.0018 \\
53469	&	2000 AX8	&	12.39	&	18.453 $\pm$ 0.354	&	5.30491 $\pm$ 0.00074	&	0.04462 $\pm$ 0.00015	&	7.4299 $\pm$ 0.0029 \\
65150	&	2002 CA126	&	12.47	&		&	5.31311 $\pm$ 0.00198	&	0.04952 $\pm$ 0.00086	&	7.4866 $\pm$ 0.0392 \\
65225	&	2002 EK44	&	12.36	&	16.654 $\pm$ 0.234	&	5.28639 $\pm$ 0.00032	&	0.04005 $\pm$ 0.00017	&	7.3903 $\pm$ 0.0031 \\
88229	&	2001 BZ54	&	12.32	&		&	5.30633 $\pm$ 0.00041	&	0.05739 $\pm$ 0.00074	&	7.4842 $\pm$ 0.0022 \\
89918	&	2002 ER33	&	12.8	&	12.373 $\pm$ 1.458	&	5.30353 $\pm$ 0.00068	&	0.03826 $\pm$ 0.00012	&	7.0953 $\pm$ 0.0015 \\
101405	&	1998 VJ3	&	13.11	&		&	5.28464 $\pm$ 0.00060	&	0.04445 $\pm$ 0.00028	&	7.4773 $\pm$ 0.0027 \\
111805	&	2002 CZ256	&	12.58	&		&	5.31320 $\pm$ 0.00192	&	0.04870 $\pm$ 0.00038	&	7.3476 $\pm$ 0.0488 \\
127846	&	2003 FO111	&	12.47	&		&	5.28895 $\pm$ 0.00037	&	0.04116 $\pm$ 0.00009	&	7.5483 $\pm$ 0.0148 \\
160661	&	1999 XD225	&	13.21	&	13.088 $\pm$ 0.703	&	5.28393 $\pm$ 0.00037	&	0.04216 $\pm$ 0.00018	&	7.5451 $\pm$ 0.0021 \\
160856	&	2001 DU92	&	12.59	&	16.216 $\pm$ 0.54	&	5.29276 $\pm$ 0.00067	&	0.06018 $\pm$ 0.00168	&	7.3411 $\pm$ 0.0011 \\
163135	&	2002 CT22	&	12.58	&	16.661 $\pm$ 0.735	&	5.28809 $\pm$ 0.00035	&	0.03962 $\pm$ 0.00011	&	7.4705 $\pm$ 0.0088 \\
163189	&	2002 EU6	&	12.9	&	16.23 $\pm$ 0.781	&	5.30304 $\pm$ 0.00063	&	0.05173 $\pm$ 0.00027	&	7.3682 $\pm$ 0.0018 \\
163216	&	2002 EN68	&	12.55	&	13.25 $\pm$ 0.801	&	5.29892 $\pm$ 0.00051	&	0.03974 $\pm$ 0.00008	&	7.2455 $\pm$ 0.0029 \\
166211	&	2002 EP135	&	12.88	&	14.412 $\pm$ 1.052	&	5.29264 $\pm$ 0.00010	&	0.04550 $\pm$ 0.00004	&	7.4303 $\pm$ 0.0005 \\
191088	&	2002 CP286	&	12.95	&		&	5.31470 $\pm$ 0.00274	&	0.05000 $\pm$ 0.00098	&	7.4788 $\pm$ 0.0463 \\
191116	&	2002 ES84	&	13.1	&	13.023 $\pm$ 1.049	&	5.28790 $\pm$ 0.00054	&	0.06098 $\pm$ 0.00022	&	7.4879 $\pm$ 0.0018 \\
192388	&	1996 RD29	&	12.91	&		&	5.29522 $\pm$ 0.00091	&	0.04449 $\pm$ 0.00016	&	7.3439 $\pm$ 0.0008 \\
192929	&	2000 AT44	&	12.5	&	13.339 $\pm$ 0.482	&	5.29852 $\pm$ 0.00087	&	0.04532 $\pm$ 0.00028	&	7.3884 $\pm$ 0.0011 \\
	&		&		&		&		&	continuation on	&	 next page 
\end{tabular}
\end{table}

\begin{table}[]
\begin{tabular}{rllcccc}
nr	&	prov. des.	&	H$^{[1]}$	&	D [km]$^{[2]}$	&	$a_{prop}$ [au]$^{[3]}$	&	$e_{prop}$$^{[3]}$	&	$i_{prop}$ [$^{\circ}$]$^{[3]}$\\
195287	&	2002 EV79	&	13.32	&	12.908 $\pm$ 1.52	&	5.29829 $\pm$ 0.00154	&	0.05352 $\pm$ 0.00029	&	7.4974 $\pm$ 0.0016 \\
195412	&	2002 GF39	&	12.45	&	19.051 $\pm$ 0.439	&	5.31429 $\pm$ 0.00208	&	0.04194 $\pm$ 0.00049	&	7.3980 $\pm$ 0.0437 \\
200024	&	2007 OO7	&	12.8	&	13.808 $\pm$ 0.886	&	5.29731 $\pm$ 0.00107	&	0.04671 $\pm$ 0.00017	&	7.3943 $\pm$ 0.0007 \\
200032	&	2007 PU43	&	12.89	&	17.945 $\pm$ 0.582	&	5.30250 $\pm$ 0.00101	&	0.04055 $\pm$ 0.00034	&	7.4355 $\pm$ 0.0008 \\
210237	&	2007 RQ154	&	12.75	&	16.698 $\pm$ 0.713	&	5.31388 $\pm$ 0.00169	&	0.04575 $\pm$ 0.00091	&	7.4411 $\pm$ 0.0493 \\
214376	&	2005 LF20	&	13.06	&		&	5.28327 $\pm$ 0.00027	&	0.04786 $\pm$ 0.00012	&	7.3828 $\pm$ 0.0012 \\
219835	&	2002 CH82	&	13.29	&		&	5.29537 $\pm$ 0.00064	&	0.06259 $\pm$ 0.00012	&	7.2826 $\pm$ 0.0011 \\
221786	&	2007 VA8	&	13.3	&		&	5.28067 $\pm$ 0.00024	&	0.04324 $\pm$ 0.00005	&	7.4014 $\pm$ 0.0011 \\
222861	&	2002 EZ134	&	12.76	&	13.004 $\pm$ 0.896	&	5.28173 $\pm$ 0.00037	&	0.05655 $\pm$ 0.00008	&	7.2876 $\pm$ 0.0021 \\
223251	&	2003 FB70	&	12.6	&	17.702 $\pm$ 0.529	&	5.28642 $\pm$ 0.00080	&	0.04997 $\pm$ 0.00010	&	7.3940 $\pm$ 0.0125 \\
225214	&	2008 RM58	&	13.5	&	11.248 $\pm$ 0.758	&	5.29248 $\pm$ 0.00046	&	0.06430 $\pm$ 0.00023	&	7.3509 $\pm$ 0.0087 \\
225222	&	2008 SP32	&	13.2	&	13.945 $\pm$ 0.652	&	5.30024 $\pm$ 0.00128	&	0.05227 $\pm$ 0.00017	&	7.4812 $\pm$ 0.0066 \\
225227	&	2008 TO65	&	12.91	&	13.241 $\pm$ 0.835	&	5.29955 $\pm$ 0.00097	&	0.03513 $\pm$ 0.00014	&	7.4048 $\pm$ 0.0083 \\
225359	&	1998 WJ24	&	13.5	&		&	5.29805 $\pm$ 0.00023	&	0.04484 $\pm$ 0.00006	&	7.2255 $\pm$ 0.0009 \\
226067	&	2002 HQ14	&	12.9	&	14.325 $\pm$ 0.812	&	5.30220 $\pm$ 0.00078	&	0.04244 $\pm$ 0.00060	&	7.4595 $\pm$ 0.0012 \\
228007	&	2007 RE27	&	13.52	&	11.085 $\pm$ 0.804	&	5.31435 $\pm$ 0.00191	&	0.04426 $\pm$ 0.00035	&	7.4952 $\pm$ 0.0391 \\
228097	&	2008 SQ31	&	13.56	&	10.481 $\pm$ 0.819	&	5.29949 $\pm$ 0.00076	&	0.05814 $\pm$ 0.00013	&	7.4061 $\pm$ 0.0103 \\
228098	&	2008 SM38	&	13.4	&		&	5.30782 $\pm$ 0.00042	&	0.04987 $\pm$ 0.00003	&	7.0712 $\pm$ 0.0066 \\
228115	&	2008 TK76	&	13.41	&	12.813 $\pm$ 0.846	&	5.29906 $\pm$ 0.00117	&	0.04669 $\pm$ 0.00017	&	7.4664 $\pm$ 0.0012 \\
228116	&	2008 TK82	&	13.48	&	12.81 $\pm$ 0.87	&	5.31022 $\pm$ 0.00218	&	0.03567 $\pm$ 0.00043	&	7.3475 $\pm$ 0.0426 \\
229822	&	2008 TA112	&	13.38	&	13.975 $\pm$ 1.452	&	5.31046 $\pm$ 0.00229	&	0.04431 $\pm$ 0.00054	&	7.2730 $\pm$ 0.0355 \\
233683	&	2008 RG113	&	13.79	&	10.171 $\pm$ 1.05	&	5.28352 $\pm$ 0.00075	&	0.04879 $\pm$ 0.00011	&	7.2041 $\pm$ 0.0018 \\
237035	&	2008 SL91	&	13.37	&	11.96 $\pm$ 0.773	&	5.30419 $\pm$ 0.00044	&	0.04261 $\pm$ 0.00044	&	7.4308 $\pm$ 0.0013 \\
243334	&	2008 TY109	&	13.4	&	12.583 $\pm$ 1.052	&	5.30130 $\pm$ 0.00060	&	0.04274 $\pm$ 0.00007	&	7.4938 $\pm$ 0.0020 \\
246145	&	2007 PE9	&	13.21	&	12.093 $\pm$ 0.601	&	5.31229 $\pm$ 0.00160	&	0.04945 $\pm$ 0.00075	&	7.4333 $\pm$ 0.0487 \\
247409	&	2002 CF79	&	12.94	&		&	5.30139 $\pm$ 0.00057	&	0.03952 $\pm$ 0.00016	&	7.3402 $\pm$ 0.0098 \\
249247	&	2008 RV9	&	13.3	&	11.881 $\pm$ 0.689	&	5.32461 $\pm$ 0.00097	&	0.04832 $\pm$ 0.00051	&	7.5323 $\pm$ 0.0323 \\
249256	&	2008 SH38	&	13.7	&	12.063 $\pm$ 1.353	&	5.29349 $\pm$ 0.00031	&	0.03868 $\pm$ 0.00008	&	7.3525 $\pm$ 0.0005 \\
249481	&	2009 TE23	&	12.9	&	14.423 $\pm$ 0.534	&	5.28729 $\pm$ 0.00063	&	0.04362 $\pm$ 0.00011	&	7.3920 $\pm$ 0.0190 \\
252683	&	2002 AE166	&	13	&		&	5.28721 $\pm$ 0.00094	&	0.04246 $\pm$ 0.00018	&	7.3334 $\pm$ 0.0081 \\
252711	&	2002 CU152	&	13.2	&		&	5.28480 $\pm$ 0.00041	&	0.04693 $\pm$ 0.00011	&	7.2473 $\pm$ 0.0029 \\
256553	&	2007 PT4	&	13.47	&	11.315 $\pm$ 1.321	&	5.28177 $\pm$ 0.00025	&	0.06126 $\pm$ 0.00007	&	7.2267 $\pm$ 0.0157 \\
259316	&	2003 FJ38	&	13.9	&		&	5.29803 $\pm$ 0.00099	&	0.05018 $\pm$ 0.00018	&	7.4150 $\pm$ 0.0008 \\
263822	&	2008 SO49	&	13.6	&	12.842 $\pm$ 1.575	&	5.28980 $\pm$ 0.00074	&	0.05301 $\pm$ 0.00029	&	7.2791 $\pm$ 0.0011 \\
263829	&	2008 SL222	&	13.7	&		&	5.31818 $\pm$ 0.00090	&	0.04803 $\pm$ 0.00029	&	7.4725 $\pm$ 0.0113 \\
263833	&	2008 TJ53	&	13.39	&		&	5.28277 $\pm$ 0.00066	&	0.04664 $\pm$ 0.00027	&	7.2566 $\pm$ 0.0015 \\
264055	&	2009 SY5	&	13.9	&		&	5.30242 $\pm$ 0.00099	&	0.05524 $\pm$ 0.00019	&	7.5283 $\pm$ 0.0015 \\
264071	&	2009 SW159	&	13.86	&	9.308 $\pm$ 0.786	&	5.30097 $\pm$ 0.00074	&	0.03994 $\pm$ 0.00020	&	7.3550 $\pm$ 0.0083 \\
264101	&	2009 SN302	&	13.57	&		&	5.29128 $\pm$ 0.00011	&	0.04558 $\pm$ 0.00162	&	7.4393 $\pm$ 0.0010 \\
264125	&	2009 TE35	&	12.87	&	14.906 $\pm$ 0.737	&	5.28883 $\pm$ 0.00059	&	0.05464 $\pm$ 0.00013	&	7.4805 $\pm$ 0.0067 \\
264139	&	2009 UP78	&	13.4	&	14.172 $\pm$ 1.54	&	5.28924 $\pm$ 0.00115	&	0.05337 $\pm$ 0.00015	&	7.3813 $\pm$ 0.0017 \\
264156	&	2009 WV5	&	13.2	&	14.642 $\pm$ 0.999	&	5.29478 $\pm$ 0.00033	&	0.06076 $\pm$ 0.00006	&	7.1825 $\pm$ 0.0007 \\
264164	&	2010 AV106	&	13.4	&	11.165 $\pm$ 1.086	&	5.28071 $\pm$ 0.00028	&	0.06010 $\pm$ 0.00192	&	7.4667 $\pm$ 0.0310 \\
264166	&	2010 AA123	&	13.02	&	16.478 $\pm$ 0.694	&	5.28619 $\pm$ 0.00071	&	0.04306 $\pm$ 0.00007	&	7.5722 $\pm$ 0.0141 \\
266647	&	2008 SH229	&	13.56	&	10.752 $\pm$ 0.956	&	5.29690 $\pm$ 0.00049	&	0.05330 $\pm$ 0.00011	&	7.3024 $\pm$ 0.0024 \\
266808	&	2009 SN355	&	13.6	&		&	5.29648 $\pm$ 0.00077	&	0.04771 $\pm$ 0.00008	&	7.3587 $\pm$ 0.0015 \\
269327	&	2008 SC278	&	13.48	&	10.273 $\pm$ 0.943	&	5.29183 $\pm$ 0.00011	&	0.04481 $\pm$ 0.00127	&	7.6236 $\pm$ 0.0011 \\
274675	&	2008 UZ7	&	13.3	&	11.657 $\pm$ 1.162	&	5.29177 $\pm$ 0.00029	&	0.05445 $\pm$ 0.00175	&	7.3320 $\pm$ 0.0005 \\
	&		&		&		&		&	continuation on	&	 next page 
\end{tabular}
\end{table}

\begin{table}[]
\begin{tabular}{rllcccc}
nr	&	prov. des.	&	H$^{[1]}$	&	D [km]$^{[2]}$	&	$a_{prop}$ [au]$^{[3]}$	&	$e_{prop}$$^{[3]}$	&	$i_{prop}$ [$^{\circ}$]$^{[3]}$\\
275116	&	2009 VD57	&	13.07	&	14.177 $\pm$ 0.73	&	5.28643 $\pm$ 0.00047	&	0.04272 $\pm$ 0.00010	&	7.4034 $\pm$ 0.0088 \\
287577	&	2003 FE42	&	13.72	&		&	5.27778 $\pm$ 0.00006	&	0.05039 $\pm$ 0.00047	&	7.3046 $\pm$ 0.0010 \\
295491	&	2008 RO20	&	14.07	&		&	5.29607 $\pm$ 0.00068	&	0.04676 $\pm$ 0.00012	&	7.5852 $\pm$ 0.0093 \\
295500	&	2008 RQ45	&	14	&		&	5.28914 $\pm$ 0.00074	&	0.04246 $\pm$ 0.00044	&	7.3525 $\pm$ 0.0028 \\
295503	&	2008 RE55	&	13.8	&		&	5.29274 $\pm$ 0.00007	&	0.04539 $\pm$ 0.00004	&	7.3813 $\pm$ 0.0006 \\
295513	&	2008 RR82	&	14	&		&	5.28740 $\pm$ 0.00069	&	0.04094 $\pm$ 0.00015	&	7.4066 $\pm$ 0.0082 \\
295520	&	2008 RP112	&	13.9	&	11.431 $\pm$ 1.236	&	5.27764 $\pm$ 0.00032	&	0.06030 $\pm$ 0.00010	&	7.5686 $\pm$ 0.0091 \\
295625	&	2008 SM232	&	13.49	&		&	5.28435 $\pm$ 0.00022	&	0.03652 $\pm$ 0.00013	&	7.2870 $\pm$ 0.0011 \\
295653	&	2008 TX4	&	13.2	&	11.381 $\pm$ 0.823	&	5.28964 $\pm$ 0.00058	&	0.05222 $\pm$ 0.00016	&	7.2854 $\pm$ 0.0012 \\
295701	&	2008 TW174	&	13.28	&	15.821 $\pm$ 1.036	&	5.31444 $\pm$ 0.00302	&	0.04917 $\pm$ 0.00063	&	7.0421 $\pm$ 0.0288 \\
296581	&	2009 RU8	&	13.69	&		&	5.28744 $\pm$ 0.00072	&	0.05135 $\pm$ 0.00013	&	7.4714 $\pm$ 0.0129 \\
296604	&	2009 RT63	&	13.7	&		&	5.27814 $\pm$ 0.00025	&	0.04696 $\pm$ 0.00024	&	7.4235 $\pm$ 0.0015 \\
306753	&	2000 YM25	&	13.58	&		&	5.29581 $\pm$ 0.00062	&	0.04419 $\pm$ 0.00012	&	7.3954 $\pm$ 0.0040 \\
312461	&	2008 RK14	&	13.4	&	13.519 $\pm$ 1.195	&	5.29706 $\pm$ 0.00064	&	0.04519 $\pm$ 0.00008	&	7.4834 $\pm$ 0.0016 \\
312620	&	2009 SA252	&	14.1	&		&	5.31502 $\pm$ 0.00190	&	0.05143 $\pm$ 0.00098	&	7.4341 $\pm$ 0.0386 \\
312622	&	2009 SH313	&	13.5	&		&	5.29419 $\pm$ 0.00046	&	0.04565 $\pm$ 0.00007	&	7.4752 $\pm$ 0.0006 \\
313024	&	2000 AV210	&	13.6	&	10.805 $\pm$ 0.599	&	5.28338 $\pm$ 0.00020	&	0.04545 $\pm$ 0.00004	&	7.4914 $\pm$ 0.0010 \\
313323	&	2002 EJ152	&	13.66	&	10.133 $\pm$ 0.548	&	5.32028 $\pm$ 0.00016	&	0.05237 $\pm$ 0.00011	&	7.5301 $\pm$ 0.0050 \\
315205	&	2007 QO15	&	13.6	&		&	5.31483 $\pm$ 0.00271	&	0.04237 $\pm$ 0.00018	&	7.4526 $\pm$ 0.0167 \\
315915	&	2008 RB123	&	13.8	&		&	5.31239 $\pm$ 0.00195	&	0.04545 $\pm$ 0.00087	&	7.2461 $\pm$ 0.0689 \\
315918	&	2008 RT126	&	14.04	&		&	5.28075 $\pm$ 0.00043	&	0.04211 $\pm$ 0.00007	&	7.2664 $\pm$ 0.0014 \\
315925	&	2008 SE96	&	13.7	&		&	5.30192 $\pm$ 0.00135	&	0.06017 $\pm$ 0.00012	&	7.3595 $\pm$ 0.0074 \\
315949	&	2008 TJ126	&	13.2	&		&	5.30950 $\pm$ 0.00252	&	0.03241 $\pm$ 0.00026	&	7.3091 $\pm$ 0.0470 \\
315950	&	2008 TT127	&	13.76	&		&	5.29796 $\pm$ 0.00087	&	0.04443 $\pm$ 0.00014	&	7.3192 $\pm$ 0.0011 \\
315954	&	2008 TJ176	&	13.76	&		&	5.31356 $\pm$ 0.00228	&	0.03449 $\pm$ 0.00043	&	7.3430 $\pm$ 0.0433 \\
316130	&	2009 RA74	&	13.4	&	10.419 $\pm$ 0.714	&	5.28197 $\pm$ 0.00050	&	0.04700 $\pm$ 0.00008	&	7.3984 $\pm$ 0.0013 \\
316157	&	2009 UT13	&	13.9	&		&	5.30583 $\pm$ 0.00019	&	0.03756 $\pm$ 0.00004	&	7.3313 $\pm$ 0.0016 \\
316165	&	2009 VP110	&	13.2	&	13.8 $\pm$ 1.144	&	5.29951 $\pm$ 0.00104	&	0.06556 $\pm$ 0.00021	&	7.4262 $\pm$ 0.0103 \\
316174	&	2009 WM250	&	13.4	&		&	5.29157 $\pm$ 0.00103	&	0.05899 $\pm$ 0.00075	&	7.2337 $\pm$ 0.0008 \\
316267	&	2010 PW25	&	13.27	&		&	5.29297 $\pm$ 0.00008	&	0.05118 $\pm$ 0.00013	&	7.4478 $\pm$ 0.0006 \\
316446	&	2010 UT53	&	13.8	&		&	5.31691 $\pm$ 0.00099	&	0.04948 $\pm$ 0.00017	&	7.4697 $\pm$ 0.0142 \\
316484	&	2010 VM61	&	13.02	&		&	5.30616 $\pm$ 0.00016	&	0.04040 $\pm$ 0.00008	&	7.3078 $\pm$ 0.0023 \\
316551	&	2010 XA84	&	13.8	&		&	5.29558 $\pm$ 0.00090	&	0.04624 $\pm$ 0.00010	&	7.3262 $\pm$ 0.0010 \\
316552	&	2010 XF87	&	13.5	&		&	5.28917 $\pm$ 0.00154	&	0.05556 $\pm$ 0.00034	&	7.5163 $\pm$ 0.0060 \\
321095	&	2008 SA277	&	13.9	&		&	5.28049 $\pm$ 0.00018	&	0.04834 $\pm$ 0.00006	&	7.2340 $\pm$ 0.0053 \\
321113	&	2008 TQ131	&	13.8	&	11.395 $\pm$ 1.525	&	5.31471 $\pm$ 0.00253	&	0.04854 $\pm$ 0.00080	&	7.3964 $\pm$ 0.0491 \\
321115	&	2008 TB142	&	13.7	&	10.609 $\pm$ 0.927	&	5.29233 $\pm$ 0.00041	&	0.05634 $\pm$ 0.00091	&	7.2441 $\pm$ 0.0011 \\
321651	&	2010 BY9	&	13.5	&		&	5.29149 $\pm$ 0.00035	&	0.05112 $\pm$ 0.00032	&	7.4820 $\pm$ 0.0006 \\
322540	&	2011 YR29	&	13.46	&		&	5.28473 $\pm$ 0.00024	&	0.04368 $\pm$ 0.00009	&	7.2906 $\pm$ 0.0015 \\
325695	&	2009 UF23	&	13.29	&	11.631 $\pm$ 1.246	&	5.28760 $\pm$ 0.00030	&	0.03490 $\pm$ 0.00012	&	7.3411 $\pm$ 0.0035 \\
328381	&	2008 RK41	&	14	&		&	5.29277 $\pm$ 0.00010	&	0.04957 $\pm$ 0.00013	&	7.4748 $\pm$ 0.0005 \\
329008	&	2010 XZ64	&	14.1	&		&	5.30289 $\pm$ 0.00043	&	0.04933 $\pm$ 0.00034	&	7.3748 $\pm$ 0.0045 \\
331050	&	2009 VH107	&	13.4	&	13.071 $\pm$ 1.317	&	5.31320 $\pm$ 0.00196	&	0.05211 $\pm$ 0.00144	&	7.3619 $\pm$ 0.0438 \\
339562	&	2005 JF162	&	13.67	&		&	5.27697 $\pm$ 0.00021	&	0.04141 $\pm$ 0.00206	&	7.8027 $\pm$ 0.0008 \\
344629	&	2003 JL18	&	13.7	&		&	5.29654 $\pm$ 0.00082	&	0.03912 $\pm$ 0.00012	&	7.3092 $\pm$ 0.0045 \\
347148	&	2010 WY	&	13.03	&		&	5.29512 $\pm$ 0.00021	&	0.05222 $\pm$ 0.00004	&	7.3093 $\pm$ 0.0005 \\
349919	&	2009 SZ166	&	14.02	&		&	5.31938 $\pm$ 0.00039	&	0.04372 $\pm$ 0.00006	&	7.3587 $\pm$ 0.0015 \\
350051	&	2010 PH25	&	13.4	&		&	5.29811 $\pm$ 0.00053	&	0.05914 $\pm$ 0.00013	&	7.3911 $\pm$ 0.0064 \\
350053	&	2010 PE49	&	13.9	&		&	5.29406 $\pm$ 0.00057	&	0.04311 $\pm$ 0.00010	&	7.3709 $\pm$ 0.0006 \\
350825	&	2002 ER38	&	13.94	&		&	5.31152 $\pm$ 0.00183	&	0.04263 $\pm$ 0.00034	&	7.3857 $\pm$ 0.0281 \\
352662	&	2008 RM14	&	13.71	&	10.434 $\pm$ 1.344	&	5.30011 $\pm$ 0.00108	&	0.05428 $\pm$ 0.00011	&	7.3620 $\pm$ 0.0016 \\
352668	&	2008 RQ72	&	13.43	&		&	5.29113 $\pm$ 0.00075	&	0.05216 $\pm$ 0.00090	&	7.2939 $\pm$ 0.0007 \\
352793	&	2008 UF189	&	14.01	&		&	5.29618 $\pm$ 0.00049	&	0.03654 $\pm$ 0.00012	&	7.3371 $\pm$ 0.0011 \\
	&		&		&		&		&	continuation on	&	 next page 
\end{tabular}
\end{table}

\begin{table}[]
\begin{tabular}{rllcccc}
nr	&	prov. des.	&	H$^{[1]}$	&	D [km]$^{[2]}$	&	$a_{prop}$ [au]$^{[3]}$	&	$e_{prop}$$^{[3]}$	&	$i_{prop}$ [$^{\circ}$]$^{[3]}$\\
353180	&	2009 RG14	&	13.8	&		&	5.29676 $\pm$ 0.00068	&	0.04648 $\pm$ 0.00014	&	7.2963 $\pm$ 0.0071 \\
353188	&	2009 RR68	&	13.8	&		&	5.31651 $\pm$ 0.00216	&	0.04966 $\pm$ 0.00059	&	7.3922 $\pm$ 0.0313 \\
353193	&	2009 SH58	&	14	&		&	5.28823 $\pm$ 0.00141	&	0.06085 $\pm$ 0.00058	&	7.4283 $\pm$ 0.0020 \\
353209	&	2009 SC354	&	13.7	&		&	5.28398 $\pm$ 0.00020	&	0.03485 $\pm$ 0.00009	&	7.1710 $\pm$ 0.0011 \\
353346	&	2010 VV78	&	14	&		&	5.29266 $\pm$ 0.00005	&	0.04259 $\pm$ 0.00002	&	7.2241 $\pm$ 0.0005 \\
353350	&	2010 VA116	&	13.81	&		&	5.29909 $\pm$ 0.00041	&	0.04136 $\pm$ 0.00009	&	7.3588 $\pm$ 0.0012 \\
353351	&	2010 VO138	&	13.6	&		&	5.27589 $\pm$ 0.00014	&	0.05045 $\pm$ 0.00334	&	7.2139 $\pm$ 0.0027 \\
353356	&	2010 VZ214	&	14.28	&		&	5.27930 $\pm$ 0.00017	&	0.04828 $\pm$ 0.00009	&	7.5331 $\pm$ 0.0012 \\
353740	&	2011 YB18	&	13.8	&		&	5.30786 $\pm$ 0.00033	&	0.04839 $\pm$ 0.00014	&	7.4186 $\pm$ 0.0019 \\
353743	&	2011 YN40	&	13.4	&		&	5.28479 $\pm$ 0.00042	&	0.04311 $\pm$ 0.00008	&	7.2671 $\pm$ 0.0012 \\
353744	&	2011 YL45	&	13.7	&		&	5.29042 $\pm$ 0.00030	&	0.04851 $\pm$ 0.00008	&	7.5597 $\pm$ 0.0012 \\
355756	&	2008 QL37	&	14	&	8.864 $\pm$ 0.965	&	5.29335 $\pm$ 0.00008	&	0.05546 $\pm$ 0.00115	&	7.5500 $\pm$ 0.0010 \\
355760	&	2008 RQ10	&	13.8	&	10.436 $\pm$ 1.217	&	5.27905 $\pm$ 0.00023	&	0.06265 $\pm$ 0.00260	&	7.2633 $\pm$ 0.0286 \\
355765	&	2008 RK37	&	13.8	&	10.896 $\pm$ 1.228	&	5.28833 $\pm$ 0.00061	&	0.04196 $\pm$ 0.00037	&	7.2167 $\pm$ 0.0189 \\
355809	&	2008 SK277	&	14	&		&	5.29128 $\pm$ 0.00010	&	0.04742 $\pm$ 0.00012	&	7.6100 $\pm$ 0.0010 \\
355814	&	2008 TU37	&	14.1	&	9.472 $\pm$ 1.234	&	5.30388 $\pm$ 0.00088	&	0.04402 $\pm$ 0.00018	&	7.2709 $\pm$ 0.0012 \\
355820	&	2008 TY96	&	13.9	&		&	5.28068 $\pm$ 0.00026	&	0.04945 $\pm$ 0.00006	&	7.5335 $\pm$ 0.0035 \\
355849	&	2008 UM109	&	13.7	&		&	5.30381 $\pm$ 0.00045	&	0.04619 $\pm$ 0.00067	&	7.4654 $\pm$ 0.0016 \\
356211	&	2009 RB63	&	14	&		&	5.31729 $\pm$ 0.00061	&	0.04073 $\pm$ 0.00014	&	7.2387 $\pm$ 0.0036 \\
356222	&	2009 SF194	&	14	&		&	5.28878 $\pm$ 0.00048	&	0.04667 $\pm$ 0.00029	&	7.1184 $\pm$ 0.0025 \\
356240	&	2009 SB354	&	13.9	&		&	5.28760 $\pm$ 0.00093	&	0.05710 $\pm$ 0.00040	&	7.4233 $\pm$ 0.0024 \\
356248	&	2009 UO12	&	14.29	&		&	5.31471 $\pm$ 0.00226	&	0.04748 $\pm$ 0.00049	&	7.1598 $\pm$ 0.0443 \\
356257	&	2009 UL140	&	13.5	&	10.569 $\pm$ 0.716	&	5.30676 $\pm$ 0.00090	&	0.05552 $\pm$ 0.00035	&	7.3423 $\pm$ 0.0096 \\
356259	&	2009 UV148	&	13.8	&	10.806 $\pm$ 1.499	&	5.29962 $\pm$ 0.00108	&	0.04990 $\pm$ 0.00017	&	7.4413 $\pm$ 0.0119 \\
356270	&	2009 WJ107	&	13.8	&		&	5.28651 $\pm$ 0.00048	&	0.04214 $\pm$ 0.00010	&	7.2332 $\pm$ 0.0066 \\
356284	&	2010 CH242	&	14	&	12.435 $\pm$ 1.383	&	5.31080 $\pm$ 0.00209	&	0.04047 $\pm$ 0.00055	&	7.5234 $\pm$ 0.0418 \\
356426	&	2010 VB122	&	14.3	&		&	5.28686 $\pm$ 0.00045	&	0.04039 $\pm$ 0.00010	&	7.3568 $\pm$ 0.0034 \\
356441	&	2010 XX7	&	14.04	&		&	5.26719 $\pm$ 0.00038	&	0.04188 $\pm$ 0.00011	&	7.8568 $\pm$ 0.0021 \\
356449	&	2010 XO79	&	13.9	&		&	5.28290 $\pm$ 0.00046	&	0.04878 $\pm$ 0.00015	&	7.2260 $\pm$ 0.0019 \\
356902	&	2011 YF56	&	14.1	&		&	5.28822 $\pm$ 0.00104	&	0.04660 $\pm$ 0.00022	&	7.4036 $\pm$ 0.0073 \\
356905	&	2011 YC75	&	14	&		&	5.28747 $\pm$ 0.00152	&	0.06048 $\pm$ 0.00026	&	7.3071 $\pm$ 0.0031 \\
356913	&	2012 BQ50	&	13.8	&		&	5.29908 $\pm$ 0.00114	&	0.05636 $\pm$ 0.00011	&	7.4266 $\pm$ 0.0018 \\
359345	&	2009 SE198	&	14.18	&		&	5.30233 $\pm$ 0.00179	&	0.04102 $\pm$ 0.00034	&	7.6295 $\pm$ 0.0081 \\
359360	&	2009 US28	&	14.3	&	9.742 $\pm$ 1.077	&	5.30179 $\pm$ 0.00100	&	0.05517 $\pm$ 0.00014	&	7.4536 $\pm$ 0.0017 \\
359361	&	2009 US55	&	14.1	&	10.084 $\pm$ 1.028	&	5.28199 $\pm$ 0.00053	&	0.05477 $\pm$ 0.00012	&	7.4286 $\pm$ 0.0024 \\
359368	&	2009 WC57	&	13.7	&	12.764 $\pm$ 0.907	&	5.28701 $\pm$ 0.00053	&	0.04424 $\pm$ 0.00014	&	7.4095 $\pm$ 0.0062 \\
359594	&	2010 VF86	&	13.9	&		&	5.28178 $\pm$ 0.00093	&	0.05059 $\pm$ 0.00005	&	7.4159 $\pm$ 0.0010 \\
359943	&	2011 YG75	&	13.9	&		&	5.29676 $\pm$ 0.00105	&	0.04762 $\pm$ 0.00018	&	7.3808 $\pm$ 0.0010 \\
360031	&	2013 AA30	&	13.4	&		&	5.28629 $\pm$ 0.00052	&	0.04119 $\pm$ 0.00014	&	7.2746 $\pm$ 0.0125 \\
360072	&	2013 AJ131	&	13.59	&		&	5.28560 $\pm$ 0.00074	&	0.05403 $\pm$ 0.00019	&	7.4961 $\pm$ 0.0051 \\
360076	&	2013 AE132	&	14.1	&		&	5.31595 $\pm$ 0.00174	&	0.04672 $\pm$ 0.00122	&	7.3339 $\pm$ 0.0497 \\
360079	&	2013 BM1	&	13.96	&		&	5.28989 $\pm$ 0.00116	&	0.05861 $\pm$ 0.00027	&	7.2547 $\pm$ 0.0010 \\
361999	&	2008 TV96	&	13.8	&	9.092 $\pm$ 1.079	&	5.29518 $\pm$ 0.00055	&	0.05280 $\pm$ 0.00007	&	7.4544 $\pm$ 0.0009 \\
365037	&	2008 SQ261	&	13.8	&		&	5.30437 $\pm$ 0.00060	&	0.04333 $\pm$ 0.00025	&	7.4032 $\pm$ 0.0053 \\
366254	&	2012 YY2	&	14	&		&	5.29325 $\pm$ 0.00017	&	0.04472 $\pm$ 0.00003	&	7.1988 $\pm$ 0.0003 \\
366317	&	2013 CV197	&	14.3	&		&	5.28734 $\pm$ 0.00049	&	0.06332 $\pm$ 0.00058	&	7.3557 $\pm$ 0.0041 \\
386967	&	2011 YS34	&	14.1	&		&	5.30048 $\pm$ 0.00092	&	0.05866 $\pm$ 0.00016	&	7.4433 $\pm$ 0.0017 \\
387390	&	2013 AG133	&	13.85	&		&	5.30063 $\pm$ 0.00085	&	0.05499 $\pm$ 0.00014	&	7.5039 $\pm$ 0.0021 \\
388891	&	2008 RW128	&	14.1	&		&	5.28979 $\pm$ 0.00039	&	0.04778 $\pm$ 0.00008	&	7.5998 $\pm$ 0.0014 \\
389317	&	2009 ST140	&	13.81	&		&	5.29662 $\pm$ 0.00034	&	0.04968 $\pm$ 0.00016	&	7.3088 $\pm$ 0.0016 \\
389318	&	2009 SV169	&	13.69	&		&	5.30732 $\pm$ 0.00050	&	0.05871 $\pm$ 0.00020	&	7.5124 $\pm$ 0.0050 \\
389327	&	2009 SB248	&	14.1	&		&	5.29868 $\pm$ 0.00081	&	0.05169 $\pm$ 0.00010	&	7.5249 $\pm$ 0.0080 \\
	&		&		&		&		&	continuation on	&	 next page 
\end{tabular}
\end{table}

\begin{table}[]
\begin{tabular}{rllcccc}
nr	&	prov. des.	&	H$^{[1]}$	&	D [km]$^{[2]}$	&	$a_{prop}$ [au]$^{[3]}$	&	$e_{prop}$$^{[3]}$	&	$i_{prop}$ [$^{\circ}$]$^{[3]}$\\
389332	&	2009 SF283	&	14	&		&	5.31827 $\pm$ 0.00031	&	0.03955 $\pm$ 0.00009	&	7.5303 $\pm$ 0.0012 \\
389560	&	2010 UN93	&	14.41	&		&	5.27717 $\pm$ 0.00020	&	0.04674 $\pm$ 0.00023	&	7.2522 $\pm$ 0.0008 \\
389823	&	2011 YP24	&	13.6	&		&	5.29054 $\pm$ 0.00099	&	0.05515 $\pm$ 0.00059	&	7.3108 $\pm$ 0.0011 \\
389824	&	2011 YH75	&	13.8	&		&	5.28569 $\pm$ 0.00092	&	0.04548 $\pm$ 0.00011	&	7.3626 $\pm$ 0.0027 \\
390322	&	2013 BA1	&	13.66	&		&	5.29003 $\pm$ 0.00110	&	0.05025 $\pm$ 0.00038	&	7.4194 $\pm$ 0.0011 \\
390344	&	2013 CP95	&	14	&		&	5.27738 $\pm$ 0.00014	&	0.05332 $\pm$ 0.00288	&	7.3953 $\pm$ 0.0072 \\
390347	&	2013 CU111	&	13.9	&		&	5.29198 $\pm$ 0.00010	&	0.04749 $\pm$ 0.00094	&	7.5701 $\pm$ 0.0007 \\
390349	&	2013 CG140	&	13.8	&		&	5.29426 $\pm$ 0.00048	&	0.04458 $\pm$ 0.00010	&	7.5158 $\pm$ 0.0008 \\
390352	&	2013 CY173	&	14.1	&		&	5.29654 $\pm$ 0.00064	&	0.05671 $\pm$ 0.00017	&	7.2346 $\pm$ 0.0008 \\
390358	&	2013 CC207	&	14.09	&		&	5.31568 $\pm$ 0.00252	&	0.05347 $\pm$ 0.00038	&	7.2911 $\pm$ 0.0471 \\
390362	&	2013 CQ217	&	14	&		&	5.28810 $\pm$ 0.00082	&	0.04783 $\pm$ 0.00010	&	7.1941 $\pm$ 0.0208 \\
391792	&	2008 RE70	&	13.8	&		&	5.29958 $\pm$ 0.00100	&	0.04749 $\pm$ 0.00012	&	7.4788 $\pm$ 0.0016 \\
392206	&	2009 SR301	&	14.1	&		&	5.30218 $\pm$ 0.00099	&	0.04990 $\pm$ 0.00011	&	7.5211 $\pm$ 0.0021 \\
392227	&	2009 VK8	&	14.3	&		&	5.28967 $\pm$ 0.00049	&	0.04059 $\pm$ 0.00067	&	7.1532 $\pm$ 0.0057 \\
392239	&	2009 WP12	&	14.1	&		&	5.30136 $\pm$ 0.00221	&	0.05854 $\pm$ 0.00059	&	7.3537 $\pm$ 0.0056 \\
392309	&	2010 CO226	&	13.8	&		&	5.29454 $\pm$ 0.00070	&	0.04672 $\pm$ 0.00009	&	7.4336 $\pm$ 0.0007 \\
392460	&	2010 XY9	&	14.2	&		&	5.28852 $\pm$ 0.00082	&	0.05463 $\pm$ 0.00016	&	7.1938 $\pm$ 0.0013 \\
392461	&	2010 XT65	&	13.8	&		&	5.30296 $\pm$ 0.00075	&	0.04292 $\pm$ 0.00081	&	7.3692 $\pm$ 0.0054 \\
392462	&	2010 XB77	&	14	&		&	5.28677 $\pm$ 0.00038	&	0.04370 $\pm$ 0.00023	&	7.4404 $\pm$ 0.0160 \\
392703	&	2011 YD71	&	13.99	&		&	5.30436 $\pm$ 0.00020	&	0.04431 $\pm$ 0.00023	&	7.5656 $\pm$ 0.0026 \\
393144	&	2013 BT60	&	13.9	&		&	5.27854 $\pm$ 0.00022	&	0.03530 $\pm$ 0.00003	&	7.2506 $\pm$ 0.0006 \\
396140	&	2013 DU5	&	14.3	&		&	5.29170 $\pm$ 0.00114	&	0.05931 $\pm$ 0.00163	&	7.0662 $\pm$ 0.0066 \\
396159	&	2013 EA41	&	14.2	&		&	5.28863 $\pm$ 0.00059	&	0.06009 $\pm$ 0.00013	&	7.2481 $\pm$ 0.0016 \\
398616	&	2011 YB3	&	13.76	&		&	5.31244 $\pm$ 0.00179	&	0.04938 $\pm$ 0.00096	&	7.3522 $\pm$ 0.0659 \\
398800	&	2013 BZ	&	13.6	&		&	5.28664 $\pm$ 0.00044	&	0.04249 $\pm$ 0.00017	&	7.3071 $\pm$ 0.0077 \\
412432	&	2014 EG31	&	13.8	&		&	5.31418 $\pm$ 0.00148	&	0.05128 $\pm$ 0.00075	&	7.3086 $\pm$ 0.0429 \\
426992	&	2014 DQ67	&	13.9	&		&	5.27808 $\pm$ 0.00014	&	0.04964 $\pm$ 0.00024	&	7.3384 $\pm$ 0.0015 \\
429083	&	2009 RP8	&	13.88	&		&	5.29211 $\pm$ 0.00034	&	0.05797 $\pm$ 0.00104	&	7.2907 $\pm$ 0.0012 \\
429974	&	2013 BM26	&	14.13	&		&	5.30657 $\pm$ 0.00035	&	0.04039 $\pm$ 0.00004	&	7.4304 $\pm$ 0.0017 \\
432264	&	2009 SZ27	&	13.8	&		&	5.31168 $\pm$ 0.00248	&	0.04355 $\pm$ 0.00041	&	7.4915 $\pm$ 0.0293 \\
432629	&	2010 VC94	&	14	&		&	5.29806 $\pm$ 0.00158	&	0.04914 $\pm$ 0.00017	&	7.3217 $\pm$ 0.0014 \\
433275	&	2013 AT93	&	13.4	&		&	5.27857 $\pm$ 0.00038	&	0.05520 $\pm$ 0.00288	&	7.5284 $\pm$ 0.0026 \\
433277	&	2013 AT108	&	14.26	&		&	5.31241 $\pm$ 0.00190	&	0.04306 $\pm$ 0.00070	&	7.5192 $\pm$ 0.0491 \\
433287	&	2013 BU44	&	14.34	&		&	5.31429 $\pm$ 0.00217	&	0.04826 $\pm$ 0.00062	&	7.4642 $\pm$ 0.0360 \\
433295	&	2013 CV209	&	14.1	&		&	5.28100 $\pm$ 0.00017	&	0.04283 $\pm$ 0.00011	&	7.8006 $\pm$ 0.0010 \\
433297	&	2013 DE13	&	14.1	&		&	5.28819 $\pm$ 0.00101	&	0.04512 $\pm$ 0.00014	&	7.3759 $\pm$ 0.0111 \\
433671	&	2014 DF32	&	14.1	&		&	5.29003 $\pm$ 0.00055	&	0.05379 $\pm$ 0.00011	&	7.3922 $\pm$ 0.0019 \\
433674	&	2014 DH124	&	14.3	&		&	5.30483 $\pm$ 0.00051	&	0.03978 $\pm$ 0.00008	&	7.6044 $\pm$ 0.0049 \\
435573	&	2008 RA82	&	14.3	&		&	5.29164 $\pm$ 0.00006	&	0.03941 $\pm$ 0.00005	&	7.5508 $\pm$ 0.0011 \\
435632	&	2008 SB137	&	14	&		&	5.29411 $\pm$ 0.00014	&	0.05613 $\pm$ 0.00038	&	7.5447 $\pm$ 0.0005 \\
436061	&	2009 RR64	&	14.4	&		&	5.29293 $\pm$ 0.00017	&	0.05436 $\pm$ 0.00121	&	7.5227 $\pm$ 0.0011 \\
436072	&	2009 SA98	&	13.9	&		&	5.30690 $\pm$ 0.00114	&	0.05462 $\pm$ 0.00032	&	7.5325 $\pm$ 0.0033 \\
436107	&	2009 SR319	&	14.2	&		&	5.28011 $\pm$ 0.00042	&	0.04347 $\pm$ 0.00019	&	7.7064 $\pm$ 0.0010 \\
436124	&	2009 TZ28	&	14.3	&		&	5.29636 $\pm$ 0.00062	&	0.04163 $\pm$ 0.00018	&	7.4197 $\pm$ 0.0042 \\
436409	&	2010 XW81	&	14.1	&		&	5.29320 $\pm$ 0.00027	&	0.04243 $\pm$ 0.00004	&	7.6353 $\pm$ 0.0007 \\
436760	&	2012 BW68	&	13.9	&		&	5.31593 $\pm$ 0.00150	&	0.04952 $\pm$ 0.00068	&	7.1467 $\pm$ 0.0422 \\
437310	&	2013 CN206	&	14.4	&		&	5.28275 $\pm$ 0.00013	&	0.03778 $\pm$ 0.00007	&	7.3191 $\pm$ 0.0030 \\
437311	&	2013 CF217	&	14	&		&	5.29774 $\pm$ 0.00036	&	0.04393 $\pm$ 0.00009	&	7.3502 $\pm$ 0.0081 \\
437720	&	2014 DB110	&	14	&		&	5.28538 $\pm$ 0.00062	&	0.06135 $\pm$ 0.00038	&	7.4157 $\pm$ 0.0034 \\
466239	&	2013 BU16	&	14.3	&		&	5.30609 $\pm$ 0.00030	&	0.04598 $\pm$ 0.00005	&	7.4806 $\pm$ 0.0021 \\
467635	&	2008 RV54	&	14.2	&		&	5.29488 $\pm$ 0.00022	&	0.05657 $\pm$ 0.00002	&	7.2476 $\pm$ 0.0004 \\
468015	&	2013 BB54	&	14.17	&		&	5.30783 $\pm$ 0.00056	&	0.04382 $\pm$ 0.00004	&	7.4237 $\pm$ 0.0057 \\
469163	&	2015 HD140	&	14.5	&		&	5.27488 $\pm$ 0.00014	&	0.03784 $\pm$ 0.00226	&	7.3685 $\pm$ 0.0006 \\
489949	&	2008 RS112	&	14.4	&		&	5.29748 $\pm$ 0.00073	&	0.05043 $\pm$ 0.00014	&	7.3571 $\pm$ 0.0119 \\
	&		&		&		&		&	continuation on	&	 next page 
\end{tabular}
\end{table}

\begin{table}[]
\begin{tabular}{rllcccc}
nr	&	prov. des.	&	H$^{[1]}$	&	D [km]$^{[2]}$	&	$a_{prop}$ [au]$^{[3]}$	&	$e_{prop}$$^{[3]}$	&	$i_{prop}$ [$^{\circ}$]$^{[3]}$\\
490809	&	2010 VE115	&	14.3	&		&	5.28136 $\pm$ 0.00044	&	0.04804 $\pm$ 0.00012	&	7.2990 $\pm$ 0.0013 \\
490817	&	2010 VF142	&	14.1	&		&	5.30254 $\pm$ 0.00071	&	0.04120 $\pm$ 0.00012	&	7.6373 $\pm$ 0.0027 \\
490820	&	2010 VA148	&	14.6	&		&	5.29666 $\pm$ 0.00072	&	0.04077 $\pm$ 0.00013	&	7.3406 $\pm$ 0.0096 \\
490839	&	2010 WD33	&	14.4	&		&	5.29024 $\pm$ 0.00012	&	0.04269 $\pm$ 0.00008	&	7.5648 $\pm$ 0.0021 \\
491834	&	2013 AQ39	&	14.2	&		&	5.28135 $\pm$ 0.00044	&	0.04558 $\pm$ 0.00012	&	7.4938 $\pm$ 0.0019 \\
491937	&	2013 CJ112	&	13.8	&		&	5.29918 $\pm$ 0.00058	&	0.04790 $\pm$ 0.00015	&	7.3347 $\pm$ 0.0009 \\
493531	&	2015 FF112	&	13.8	&		&	5.29339 $\pm$ 0.00025	&	0.04621 $\pm$ 0.00005	&	7.3987 $\pm$ 0.0009 \\
493534	&	2015 FU206	&	14.4	&		&	5.30315 $\pm$ 0.00108	&	0.04944 $\pm$ 0.00058	&	7.4456 $\pm$ 0.0016 \\
495194	&	2013 BH	&	14.1	&		&	5.29243 $\pm$ 0.00020	&	0.04804 $\pm$ 0.00003	&	7.1808 $\pm$ 0.0004 \\
496031	&	2008 SR274	&	13.9	&		&	5.29794 $\pm$ 0.00060	&	0.05049 $\pm$ 0.00011	&	7.3290 $\pm$ 0.0113 \\
496300	&	2013 CV194	&	14.4	&		&	5.28224 $\pm$ 0.00032	&	0.04208 $\pm$ 0.00010	&	7.3580 $\pm$ 0.0027 \\
507354	&	2011 UQ402	&	14.47	&		&	5.30687 $\pm$ 0.00034	&	0.05115 $\pm$ 0.00015	&	7.2940 $\pm$ 0.0032 \\
507370	&	2011 YF75	&	14.2	&		&	5.29222 $\pm$ 0.00029	&	0.04961 $\pm$ 0.00009	&	7.1441 $\pm$ 0.0008 \\
510519	&	2012 BB155	&	14.1	&		&	5.31695 $\pm$ 0.00119	&	0.04861 $\pm$ 0.00042	&	7.1741 $\pm$ 0.0400 \\
542169	&	2013 AD8	&	13.57	&		&	5.30997 $\pm$ 0.00214	&	0.04463 $\pm$ 0.00050	&	7.3777 $\pm$ 0.0201 \\
542262	&	2013 BL	&	13.7	&		&	5.28851 $\pm$ 0.00085	&	0.05481 $\pm$ 0.00031	&	7.2794 $\pm$ 0.0011 \\
542275	&	2013 BY16	&	13.88	&		&	5.30581 $\pm$ 0.00036	&	0.04280 $\pm$ 0.00004	&	7.3603 $\pm$ 0.0031 \\
542323	&	2013 CR12	&	14.2	&		&	5.31364 $\pm$ 0.00221	&	0.05294 $\pm$ 0.00082	&	7.3147 $\pm$ 0.0535 \\
542399	&	2013 CH95	&	13.9	&		&	5.28234 $\pm$ 0.00036	&	0.04884 $\pm$ 0.00021	&	7.4420 $\pm$ 0.0007 \\
546419	&	2010 VJ119	&	14.3	&		&	5.29841 $\pm$ 0.00059	&	0.04121 $\pm$ 0.00008	&	7.1403 $\pm$ 0.0062 \\
546454	&	2010 VD159	&	14.2	&		&	5.29055 $\pm$ 0.00013	&	0.03987 $\pm$ 0.00013	&	7.3569 $\pm$ 0.0018 \\
546460	&	2010 VF166	&	14.4	&		&	5.28463 $\pm$ 0.00034	&	0.05069 $\pm$ 0.00021	&	7.3203 $\pm$ 0.0068 \\
546752	&	2010 XO67	&	14.2	&		&	5.28369 $\pm$ 0.00026	&	0.04545 $\pm$ 0.00022	&	7.2955 $\pm$ 0.0014 \\
	&	2001 DD120	&	14.6	&		&	5.30468 $\pm$ 0.00042	&	0.04027 $\pm$ 0.00007	&	7.3710 $\pm$ 0.0016 \\
	&	2002 EF171	&	14.3	&		&	5.28473 $\pm$ 0.00042	&	0.05087 $\pm$ 0.00019	&	7.3110 $\pm$ 0.0208 \\
	&	2002 EN157	&	14.3	&		&	5.29465 $\pm$ 0.00013	&	0.05854 $\pm$ 0.00005	&	7.3953 $\pm$ 0.0005 \\
	&	2002 FA42	&	14.1	&		&	5.30186 $\pm$ 0.00113	&	0.03801 $\pm$ 0.00025	&	7.4180 $\pm$ 0.0021 \\
	&	2002 FJ19	&	14.5	&		&	5.30476 $\pm$ 0.00052	&	0.03945 $\pm$ 0.00008	&	7.3969 $\pm$ 0.0026 \\
	&	2002 GH197	&	14.2	&		&	5.31591 $\pm$ 0.00210	&	0.04975 $\pm$ 0.00048	&	7.3294 $\pm$ 0.0204 \\
	&	2003 DU22	&	14.4	&		&	5.30681 $\pm$ 0.00016	&	0.04334 $\pm$ 0.00003	&	7.3131 $\pm$ 0.0012 \\
	&	2003 GO64	&	14.3	&		&	5.27865 $\pm$ 0.00009	&	0.03853 $\pm$ 0.00004	&	7.3525 $\pm$ 0.0006 \\
	&	2004 GK85	&	13.7	&		&	5.30908 $\pm$ 0.00145	&	0.03367 $\pm$ 0.00034	&	7.3711 $\pm$ 0.0189 \\
	&	2005 DK4	&	14.4	&		&	5.29447 $\pm$ 0.00040	&	0.04522 $\pm$ 0.00006	&	7.3736 $\pm$ 0.0008 \\
	&	2007 RE166	&	14.5	&		&	5.31145 $\pm$ 0.00272	&	0.04864 $\pm$ 0.00079	&	7.3026 $\pm$ 0.0660 \\
	&	2008 QK42	&	14.4	&		&	5.29834 $\pm$ 0.00030	&	0.04608 $\pm$ 0.00005	&	7.4727 $\pm$ 0.0016 \\
	&	2008 RC150	&	14.2	&		&	5.31403 $\pm$ 0.00320	&	0.04545 $\pm$ 0.00060	&	7.3809 $\pm$ 0.0448 \\
	&	2008 RG65	&	14.2	&		&	5.30187 $\pm$ 0.00037	&	0.05455 $\pm$ 0.00017	&	7.5069 $\pm$ 0.0026 \\
	&	2008 RK127	&	14.1	&		&	5.29491 $\pm$ 0.00071	&	0.05326 $\pm$ 0.00010	&	7.4462 $\pm$ 0.0006 \\
	&	2008 RR161	&	13.9	&		&	5.31944 $\pm$ 0.00048	&	0.04781 $\pm$ 0.00014	&	7.4598 $\pm$ 0.0026 \\
	&	2008 RS149	&	14.2	&		&	5.30661 $\pm$ 0.00035	&	0.03531 $\pm$ 0.00010	&	7.3202 $\pm$ 0.0007 \\
	&	2008 SH232	&	14.1	&		&	5.29076 $\pm$ 0.00016	&	0.04943 $\pm$ 0.00005	&	7.5983 $\pm$ 0.0009 \\
	&	2008 SP275	&	14.4	&		&	5.31820 $\pm$ 0.00049	&	0.04420 $\pm$ 0.00015	&	7.5274 $\pm$ 0.0033 \\
	&	2008 SU275	&	14.4	&		&	5.27953 $\pm$ 0.00022	&	0.04626 $\pm$ 0.00004	&	7.2280 $\pm$ 0.0011 \\
	&	2008 SW73	&	14.1	&		&	5.28853 $\pm$ 0.00049	&	0.04498 $\pm$ 0.00011	&	7.3335 $\pm$ 0.0113 \\
	&	2008 TF194	&	14.2	&		&	5.27891 $\pm$ 0.00038	&	0.04682 $\pm$ 0.00009	&	7.2062 $\pm$ 0.0014 \\
	&	2008 TH97	&	14.4	&		&	5.29052 $\pm$ 0.00090	&	0.05304 $\pm$ 0.00020	&	7.4797 $\pm$ 0.0009 \\
	&	2008 UQ400	&	14.3	&		&	5.29320 $\pm$ 0.00010	&	0.04769 $\pm$ 0.00003	&	7.2687 $\pm$ 0.0003 \\
	&	2008 UW376	&	13.6	&		&	5.27183 $\pm$ 0.00028	&	0.04282 $\pm$ 0.00007	&	7.6491 $\pm$ 0.0032 \\
	&	2008 UY128	&	13.8	&		&	5.29182 $\pm$ 0.00018	&	0.04855 $\pm$ 0.00106	&	7.4447 $\pm$ 0.0006 \\
	&	2009 RG57	&	14.3	&		&	5.28692 $\pm$ 0.00059	&	0.05348 $\pm$ 0.00017	&	7.4748 $\pm$ 0.0071 \\
	&	2009 RM64	&	14.7	&		&	5.30844 $\pm$ 0.00085	&	0.04780 $\pm$ 0.00009	&	7.2830 $\pm$ 0.0026 \\
	&	2009 SB326	&	13.6	&		&	5.28234 $\pm$ 0.00039	&	0.05007 $\pm$ 0.00005	&	7.2500 $\pm$ 0.0012 \\
	&	2009 SF148	&	14.3	&		&	5.27451 $\pm$ 0.00016	&	0.05702 $\pm$ 0.00004	&	7.5491 $\pm$ 0.0010 \\
	&		&		&		&		&	continuation on	&	 next page 
\end{tabular}
\end{table}

\begin{table}[]
\begin{tabular}{rllcccc}
nr	&	prov. des.	&	H$^{[1]}$	&	D [km]$^{[2]}$	&	$a_{prop}$ [au]$^{[3]}$	&	$e_{prop}$$^{[3]}$	&	$i_{prop}$ [$^{\circ}$]$^{[3]}$\\
	&	2009 SF391	&	14.4	&		&	5.28771 $\pm$ 0.00116	&	0.04943 $\pm$ 0.00014	&	7.4115 $\pm$ 0.0040 \\
	&	2009 SG355	&	14.7	&		&	5.28988 $\pm$ 0.00103	&	0.05970 $\pm$ 0.00035	&	7.1440 $\pm$ 0.0009 \\
	&	2009 SJ389	&	14.1	&		&	5.28780 $\pm$ 0.00052	&	0.05348 $\pm$ 0.00016	&	7.2953 $\pm$ 0.0018 \\
	&	2009 SL48	&	14.3	&		&	5.29257 $\pm$ 0.00014	&	0.05293 $\pm$ 0.00087	&	7.4199 $\pm$ 0.0008 \\
	&	2009 SM188	&	14.3	&		&	5.28991 $\pm$ 0.00081	&	0.04962 $\pm$ 0.00019	&	7.5482 $\pm$ 0.0022 \\
	&	2009 SS301	&	14.4	&		&	5.30166 $\pm$ 0.00102	&	0.03807 $\pm$ 0.00011	&	7.6573 $\pm$ 0.0017 \\
	&	2009 SU355	&	14.4	&		&	5.28688 $\pm$ 0.00119	&	0.04673 $\pm$ 0.00010	&	7.5896 $\pm$ 0.0152 \\
	&	2009 SX354	&	14.3	&		&	5.31025 $\pm$ 0.00200	&	0.04025 $\pm$ 0.00066	&	7.4104 $\pm$ 0.0409 \\
	&	2009 SX391	&	14.6	&		&	5.28473 $\pm$ 0.00044	&	0.05196 $\pm$ 0.00016	&	7.3298 $\pm$ 0.0056 \\
	&	2009 SY373	&	13.6	&		&	5.31678 $\pm$ 0.00281	&	0.04902 $\pm$ 0.00028	&	7.3629 $\pm$ 0.0137 \\
	&	2009 TR53	&	13.7	&		&	5.31481 $\pm$ 0.00163	&	0.05976 $\pm$ 0.00101	&	7.5387 $\pm$ 0.0689 \\
	&	2009 UD168	&	14.5	&		&	5.28525 $\pm$ 0.00022	&	0.04205 $\pm$ 0.00017	&	7.3752 $\pm$ 0.0014 \\
	&	2009 UE159	&	14	&		&	5.28742 $\pm$ 0.00078	&	0.04558 $\pm$ 0.00021	&	7.3363 $\pm$ 0.0100 \\
	&	2009 UG170	&	14.3	&		&	5.30536 $\pm$ 0.00028	&	0.04409 $\pm$ 0.00013	&	7.3725 $\pm$ 0.0023 \\
	&	2009 UG57	&	14.3	&		&	5.27916 $\pm$ 0.00044	&	0.03975 $\pm$ 0.00004	&	7.2532 $\pm$ 0.0009 \\
	&	2009 VF54	&	14.6	&		&	5.29727 $\pm$ 0.00022	&	0.06547 $\pm$ 0.00004	&	7.3167 $\pm$ 0.0005 \\
	&	2009 WA142	&	14.6	&		&	5.30374 $\pm$ 0.00126	&	0.05554 $\pm$ 0.00057	&	7.4370 $\pm$ 0.0081 \\
	&	2009 WJ160	&	14.6	&		&	5.28212 $\pm$ 0.00008	&	0.03844 $\pm$ 0.00006	&	7.3186 $\pm$ 0.0011 \\
	&	2009 WO137	&	14.3	&		&	5.30489 $\pm$ 0.00090	&	0.04793 $\pm$ 0.00017	&	7.3540 $\pm$ 0.0023 \\
	&	2010 TW192	&	14.4	&		&	5.28852 $\pm$ 0.00129	&	0.05533 $\pm$ 0.00014	&	7.4029 $\pm$ 0.0016 \\
	&	2010 VF136	&	14.8	&		&	5.28704 $\pm$ 0.00033	&	0.04538 $\pm$ 0.00020	&	7.7117 $\pm$ 0.0067 \\
	&	2010 VS244	&	14.5	&		&	5.31323 $\pm$ 0.00256	&	0.04733 $\pm$ 0.00068	&	7.6120 $\pm$ 0.0400 \\
	&	2010 VZ224	&	14.2	&		&	5.28210 $\pm$ 0.00023	&	0.04333 $\pm$ 0.00010	&	7.5051 $\pm$ 0.0040 \\
	&	2010 WZ18	&	14.6	&		&	5.30468 $\pm$ 0.00034	&	0.04693 $\pm$ 0.00106	&	7.6465 $\pm$ 0.0021 \\
	&	2010 XD106	&	14.6	&		&	5.28778 $\pm$ 0.00064	&	0.05005 $\pm$ 0.00025	&	7.6303 $\pm$ 0.0046 \\
	&	2010 XU106	&	14.7	&		&	5.28342 $\pm$ 0.00047	&	0.04492 $\pm$ 0.00009	&	7.2284 $\pm$ 0.0009 \\
	&	2011 WG69	&	14	&		&	5.30854 $\pm$ 0.00036	&	0.05201 $\pm$ 0.00006	&	7.4741 $\pm$ 0.0053 \\
	&	2011 XY1	&	14.3	&		&	5.28374 $\pm$ 0.00035	&	0.05212 $\pm$ 0.00021	&	7.4192 $\pm$ 0.0042 \\
	&	2011 YM90	&	14.9	&		&	5.28670 $\pm$ 0.00127	&	0.06035 $\pm$ 0.00043	&	7.3066 $\pm$ 0.0031 \\
	&	2012 BD155	&	14.2	&		&	5.29424 $\pm$ 0.00037	&	0.03578 $\pm$ 0.00007	&	7.3295 $\pm$ 0.0010 \\
	&	2012 BU107	&	14.8	&		&	5.30454 $\pm$ 0.00041	&	0.03862 $\pm$ 0.00017	&	6.9884 $\pm$ 0.0027 \\
	&	2012 BU172	&	14.6	&		&	5.28615 $\pm$ 0.00070	&	0.04572 $\pm$ 0.00011	&	7.4598 $\pm$ 0.0067 \\
	&	2012 CT57	&	14.9	&		&	5.28805 $\pm$ 0.00088	&	0.03928 $\pm$ 0.00061	&	7.1233 $\pm$ 0.0058 \\
	&	2012 YS9	&	14.7	&		&	5.31481 $\pm$ 0.00211	&	0.04758 $\pm$ 0.00126	&	7.2766 $\pm$ 0.0310 \\
	&	2013 AJ133	&	14.2	&		&	5.30577 $\pm$ 0.00014	&	0.04198 $\pm$ 0.00005	&	7.2517 $\pm$ 0.0017 \\
	&	2013 AN129	&	13.8	&		&	5.29564 $\pm$ 0.00094	&	0.04982 $\pm$ 0.00017	&	7.4682 $\pm$ 0.0034 \\
	&	2013 AS64	&	14.3	&		&	5.28135 $\pm$ 0.00042	&	0.04573 $\pm$ 0.00007	&	7.2800 $\pm$ 0.0026 \\
	&	2013 AU35	&	13.8	&		&	5.28124 $\pm$ 0.00037	&	0.04027 $\pm$ 0.00006	&	7.5409 $\pm$ 0.0009 \\
	&	2013 AV135	&	13.8	&		&	5.31461 $\pm$ 0.00238	&	0.04957 $\pm$ 0.00068	&	7.4961 $\pm$ 0.0166 \\
	&	2013 BB17	&	14.5	&		&	5.28462 $\pm$ 0.00083	&	0.05590 $\pm$ 0.00046	&	7.2268 $\pm$ 0.0089 \\
	&	2013 BD	&	14.3	&		&	5.30519 $\pm$ 0.00019	&	0.04708 $\pm$ 0.00023	&	7.4620 $\pm$ 0.0021 \\
	&	2013 BD17	&	14.3	&		&	5.31308 $\pm$ 0.00327	&	0.03312 $\pm$ 0.00031	&	7.2749 $\pm$ 0.0234 \\
	&	2013 BE	&	14.6	&		&	5.27946 $\pm$ 0.00036	&	0.06045 $\pm$ 0.00412	&	7.5400 $\pm$ 0.0336 \\
	&	2013 BF	&	14.8	&		&	5.27734 $\pm$ 0.00018	&	0.06126 $\pm$ 0.00018	&	7.5134 $\pm$ 0.0041 \\
	&	2013 BH31	&	14.1	&		&	5.28954 $\pm$ 0.00044	&	0.04440 $\pm$ 0.00018	&	7.4044 $\pm$ 0.0038 \\
	&	2013 BJ11	&	14.2	&		&	5.29671 $\pm$ 0.00056	&	0.04226 $\pm$ 0.00012	&	7.3170 $\pm$ 0.0010 \\
	&	2013 BO82	&	14.7	&		&	5.28089 $\pm$ 0.00018	&	0.04770 $\pm$ 0.00006	&	7.1978 $\pm$ 0.0015 \\
	&	2013 BP34	&	14.1	&		&	5.29500 $\pm$ 0.00018	&	0.05958 $\pm$ 0.00003	&	7.3897 $\pm$ 0.0006 \\
	&	2013 BR1	&	14.6	&		&	5.29265 $\pm$ 0.00011	&	0.03961 $\pm$ 0.00002	&	7.3226 $\pm$ 0.0009 \\
	&	2013 BR7	&	14.2	&		&	5.30335 $\pm$ 0.00042	&	0.04797 $\pm$ 0.00024	&	7.5073 $\pm$ 0.0027 \\
	&	2013 BS38	&	14.1	&		&	5.28510 $\pm$ 0.00076	&	0.06048 $\pm$ 0.00014	&	7.3364 $\pm$ 0.0067 \\
	&	2013 BV16	&	14.1	&		&	5.29393 $\pm$ 0.00020	&	0.04694 $\pm$ 0.00003	&	7.4546 $\pm$ 0.0003 \\
	&		&		&		&		&	continuation on	&	 next page 
\end{tabular}
\end{table}

\begin{table}[]
\begin{tabular}{rllcccc}
nr	&	prov. des.	&	H$^{[1]}$	&	D [km]$^{[2]}$	&	$a_{prop}$ [au]$^{[3]}$	&	$e_{prop}$$^{[3]}$	&	$i_{prop}$ [$^{\circ}$]$^{[3]}$\\
	&	2013 BV41	&	14.4	&		&	5.30819 $\pm$ 0.00052	&	0.06525 $\pm$ 0.00028	&	7.4349 $\pm$ 0.0062 \\
	&	2013 BZ15	&	14.6	&		&	5.31549 $\pm$ 0.00176	&	0.03840 $\pm$ 0.00051	&	7.3675 $\pm$ 0.0240 \\
	&	2013 CC160	&	14.4	&		&	5.31225 $\pm$ 0.00259	&	0.03949 $\pm$ 0.00042	&	7.3507 $\pm$ 0.0322 \\
	&	2013 CF200	&	14.4	&		&	5.28209 $\pm$ 0.00031	&	0.05150 $\pm$ 0.00007	&	7.2651 $\pm$ 0.0014 \\
	&	2013 CH204	&	14.2	&		&	5.28950 $\pm$ 0.00030	&	0.03898 $\pm$ 0.00025	&	7.1040 $\pm$ 0.0095 \\
	&	2013 CM101	&	14.2	&		&	5.29647 $\pm$ 0.00063	&	0.04090 $\pm$ 0.00010	&	7.3218 $\pm$ 0.0010 \\
	&	2013 CM11	&	14.3	&		&	5.29468 $\pm$ 0.00015	&	0.06154 $\pm$ 0.00094	&	7.5909 $\pm$ 0.0007 \\
	&	2013 CM170	&	14.1	&		&	5.30052 $\pm$ 0.00044	&	0.03613 $\pm$ 0.00017	&	7.3288 $\pm$ 0.0060 \\
	&	2013 CM223	&	15.1	&		&	5.30406 $\pm$ 0.00059	&	0.05235 $\pm$ 0.00042	&	7.4547 $\pm$ 0.0014 \\
	&	2013 CN157	&	14.4	&		&	5.28020 $\pm$ 0.00049	&	0.04938 $\pm$ 0.00017	&	7.5830 $\pm$ 0.0016 \\
	&	2013 CO93	&	14.4	&		&	5.29310 $\pm$ 0.00008	&	0.05221 $\pm$ 0.00005	&	7.3486 $\pm$ 0.0007 \\
	&	2013 CR223	&	14.4	&		&	5.27789 $\pm$ 0.00015	&	0.04084 $\pm$ 0.00005	&	7.2760 $\pm$ 0.0006 \\
	&	2013 CS194	&	14.3	&		&	5.27927 $\pm$ 0.00034	&	0.05332 $\pm$ 0.00015	&	7.3158 $\pm$ 0.0083 \\
	&	2013 EL154	&	14.8	&		&	5.28408 $\pm$ 0.00054	&	0.04349 $\pm$ 0.00035	&	7.6092 $\pm$ 0.0031 \\
	&	2013 EM154	&	14.1	&		&	5.30736 $\pm$ 0.00029	&	0.05066 $\pm$ 0.00005	&	7.4626 $\pm$ 0.0020 \\
	&	2014 DM49	&	14.2	&		&	5.28727 $\pm$ 0.00066	&	0.04609 $\pm$ 0.00022	&	7.2591 $\pm$ 0.0112 \\
	&	2014 DX143	&	14.7	&		&	5.30650 $\pm$ 0.00020	&	0.04372 $\pm$ 0.00012	&	7.6318 $\pm$ 0.0003 \\
	&	2014 EG188	&	14.3	&		&	5.29268 $\pm$ 0.00015	&	0.04449 $\pm$ 0.00002	&	7.2062 $\pm$ 0.0005 \\
	&	2014 EG68	&	14.1	&		&	5.27838 $\pm$ 0.00020	&	0.06209 $\pm$ 0.00309	&	7.2843 $\pm$ 0.0347 \\
	&	2014 EK55	&	14.6	&		&	5.29908 $\pm$ 0.00106	&	0.05083 $\pm$ 0.00014	&	7.1927 $\pm$ 0.0071 \\
	&	2014 EL155	&	14.1	&		&	5.28982 $\pm$ 0.00079	&	0.04643 $\pm$ 0.00045	&	7.2772 $\pm$ 0.0021 \\
	&	2014 EM36	&	13.7	&		&	5.29154 $\pm$ 0.00020	&	0.04560 $\pm$ 0.00014	&	7.3784 $\pm$ 0.0005 \\
	&	2014 EO56	&	14.7	&		&	5.31810 $\pm$ 0.00047	&	0.04720 $\pm$ 0.00015	&	7.1782 $\pm$ 0.0048 \\
	&	2014 ES95	&	14.4	&		&	5.29502 $\pm$ 0.00060	&	0.04928 $\pm$ 0.00016	&	7.6065 $\pm$ 0.0054 \\
	&	2014 ET118	&	14.9	&		&	5.28428 $\pm$ 0.00026	&	0.04316 $\pm$ 0.00007	&	7.4103 $\pm$ 0.0015 \\
	&	2014 EV177	&	14.9	&		&	5.30861 $\pm$ 0.00035	&	0.05480 $\pm$ 0.00008	&	7.5591 $\pm$ 0.0024 \\
	&	2014 EV198	&	14.4	&		&	5.28998 $\pm$ 0.00031	&	0.03772 $\pm$ 0.00011	&	7.0962 $\pm$ 0.0064 \\
	&	2014 EV65	&	14.5	&		&	5.28611 $\pm$ 0.00018	&	0.05235 $\pm$ 0.00010	&	7.1992 $\pm$ 0.0234 \\
	&	2014 EW31	&	13.9	&		&	5.28190 $\pm$ 0.00027	&	0.04854 $\pm$ 0.00010	&	7.2329 $\pm$ 0.0017 \\
	&	2014 EX69	&	14.1	&		&	5.28533 $\pm$ 0.00028	&	0.05137 $\pm$ 0.00004	&	7.3139 $\pm$ 0.0042 \\
	&	2014 EZ194	&	14.4	&		&	5.28941 $\pm$ 0.00104	&	0.05286 $\pm$ 0.00058	&	7.2121 $\pm$ 0.0013 \\
	&	2014 EZ84	&	14.5	&		&	5.28838 $\pm$ 0.00059	&	0.04797 $\pm$ 0.00022	&	7.3654 $\pm$ 0.0100 \\
	&	2014 FD72	&	14.6	&		&	5.30554 $\pm$ 0.00059	&	0.04434 $\pm$ 0.00020	&	7.3119 $\pm$ 0.0017 \\
	&	2014 FG5	&	14.6	&		&	5.29643 $\pm$ 0.00054	&	0.05283 $\pm$ 0.00016	&	7.3145 $\pm$ 0.0030 \\
	&	2014 FT43	&	14.5	&		&	5.30899 $\pm$ 0.00107	&	0.04689 $\pm$ 0.00024	&	7.3105 $\pm$ 0.0286 \\
	&	2014 GL9	&	14.2	&		&	5.28302 $\pm$ 0.00023	&	0.04723 $\pm$ 0.00011	&	7.3426 $\pm$ 0.0013 \\
	&	2014 GM9	&	13.6	&		&	5.32418 $\pm$ 0.00232	&	0.04061 $\pm$ 0.00044	&	7.4104 $\pm$ 0.0117 \\
	&	2014 GN10	&	13.9	&		&	5.28296 $\pm$ 0.00060	&	0.04155 $\pm$ 0.00018	&	7.1739 $\pm$ 0.0014 \\
	&	2015 DT225	&	14	&		&	5.30021 $\pm$ 0.00120	&	0.05943 $\pm$ 0.00026	&	7.4643 $\pm$ 0.0117 \\
	&	2015 FF354	&	14.2	&		&	5.28313 $\pm$ 0.00031	&	0.04028 $\pm$ 0.00011	&	7.3431 $\pm$ 0.0020 \\
	&	2015 FJ139	&	14.3	&		&	5.28480 $\pm$ 0.00041	&	0.04974 $\pm$ 0.00024	&	7.5923 $\pm$ 0.0052 \\
	&	2015 FJ74	&	14.1	&		&	5.30446 $\pm$ 0.00041	&	0.04539 $\pm$ 0.00022	&	7.2931 $\pm$ 0.0034 \\
	&	2015 FQ211	&	14.4	&		&	5.28382 $\pm$ 0.00051	&	0.04879 $\pm$ 0.00056	&	7.2638 $\pm$ 0.0026 \\
	&	2015 FQ305	&	14.4	&		&	5.31609 $\pm$ 0.00206	&	0.04965 $\pm$ 0.00059	&	7.4760 $\pm$ 0.0415 \\
	&	2015 FQ357	&	14.2	&		&	5.29301 $\pm$ 0.00011	&	0.04931 $\pm$ 0.00002	&	7.2904 $\pm$ 0.0004 \\
	&	2015 FR388	&	13.9	&		&	5.30710 $\pm$ 0.00022	&	0.04775 $\pm$ 0.00003	&	7.3857 $\pm$ 0.0019 \\
	&	2015 FT170	&	14.3	&		&	5.31481 $\pm$ 0.00161	&	0.04912 $\pm$ 0.00093	&	7.1663 $\pm$ 0.0734 \\
	&	2015 HL89	&	14.1	&		&	5.27779 $\pm$ 0.00019	&	0.05750 $\pm$ 0.00306	&	7.3254 $\pm$ 0.0166 \\
	&	2015 KJ64	&	13.9	&		&	5.29173 $\pm$ 0.00014	&	0.04005 $\pm$ 0.00003	&	7.4752 $\pm$ 0.0011 \\
	&	2015 KW64	&	14.4	&		&	5.28895 $\pm$ 0.00080	&	0.04935 $\pm$ 0.00025	&	7.2849 $\pm$ 0.0038 \\
	&	2016 GD174	&	14.5	&		&	5.29823 $\pm$ 0.00050	&	0.04635 $\pm$ 0.00008	&	7.3161 $\pm$ 0.0049 \\
	&	2016 GP150	&	14.3	&		&	5.30621 $\pm$ 0.00063	&	0.05609 $\pm$ 0.00044	&	7.4113 $\pm$ 0.0044 \\
	&		&		&		&		&	continuation on	&	 next page 
\end{tabular}
\end{table}

\begin{table}[]
\begin{tabular}{rllcccc}
nr	&	prov. des.	&	H$^{[1]}$	&	D [km]$^{[2]}$	&	$a_{prop}$ [au]$^{[3]}$	&	$e_{prop}$$^{[3]}$	&	$i_{prop}$ [$^{\circ}$]$^{[3]}$\\
	&	2016 GZ187	&	14.4	&		&	5.29642 $\pm$ 0.00100	&	0.04528 $\pm$ 0.00019	&	7.3540 $\pm$ 0.0095 \\
	&	2016 HQ11	&	14	&		&	5.31741 $\pm$ 0.00217	&	0.05620 $\pm$ 0.00056	&	7.4115 $\pm$ 0.0393 \\
	&	2016 KN9	&	14.4	&		&	5.30576 $\pm$ 0.00037	&	0.04687 $\pm$ 0.00010	&	7.3294 $\pm$ 0.0024 \\
	&	2015 FM155	&	14.2	&		&	5.28549 $\pm$ 0.00046	&	0.03636 $\pm$ 0.00011	&	7.2391 $\pm$ 0.0011 

\end{tabular}
\end{table}

\begin{table}[]
\caption{Color data for Eurybates family members identified by the hierarchical clustering method (HCM) including the likely interloper (5258). The table gives the asteroid number ($\#$), the provisional designation (prov. des.), the absolute magnitude (H), the diameter (D), as well as the g-i color from the Sloan Digital Sky Survey, the spectral slopes between $0.3$ to $0.9 \mu$m (S), and the g-r color from the Zwicky Transient Facility Observations, ZTF. A machine readable version of this table is available on https://zenodo.org/SOME-LINK\\
$\left[1\right]$ Minor Planet Center; 2020-12-08, https://minorplanetcenter.net//iau/lists/JupiterTrojans.html\\
$\left[2\right]$ NEOWISE data v2.0, \cite{Mainzer2019}, https://sbn.psi.edu/pds/resource/neowisediam.html\\
$\left[4\right]$ Sloan Digital Sky Survey (SDSS) Moving Object Catalog \citep{Ivezic2001}, https://sbn.psi.edu/pds/resource/sdssmoc.html\\
$\left[5\right]$ \cite{Fornasier2007}, https://sbn.psi.edu/pds/resource/fornasier.html\\
$\left[6\right]$ \cite{SchemelBrown2021} }
\label{tab:EurybatesFamilyColour}
\begin{tabular}{rllllll}

nr	&	prov. des.	&	H$^{[1]}$	&	D [km]$^{[2]}$	&	g-i [mag]$^{[4]}$     &	S $[\%/10^3 \AA]^{[5]}$	&	g-r [mag]$^{[6]}$\\

3548	&	1973 SO	&	9.85	&	63.885	$\pm$	0.299	&				&	-0.18	$\pm$	0.57	&	0.51	$^{+	0.02	}_{-	0.02	}$ \\
5258	&	1989 AU1	&	10.33	&	53.275	$\pm$	4.429	&				&				&	0.6	$^{+	0.02	}_{-	0.02	}$ \\
8060	&	1973 SD1	&	10.95	&	37.873	$\pm$	0.567	&				&				&	0.51	$^{+	0.02	}_{-	0.02	}$ \\
9818	&	6591 P-L	&	11.07	&	28.076	$\pm$	3.215	&	0.64	$\pm$	0.022	&	2.12	$\pm$	0.72	&	0.49	$^{+	0.03	}_{-	0.04	}$ \\
13862	&	1999 XT160	&	11.6	&	24.835	$\pm$	0.589	&				&	1.59	$\pm$	0.7	&	0.46	$^{+	0.05	}_{-	0.04	}$ \\
18060	&	1999 XJ156	&	11.12	&	36.431	$\pm$	3.966	&	0.69	$\pm$	0.045	&	2.86	$\pm$	0.6	&	0.51	$^{+	0.04	}_{-	0.05	}$ \\
24380	&	2000 AA160	&	11.2	&	31.607	$\pm$	0.266	&				&	0.34	$\pm$	0.65	&	0.51	$^{+	0.02	}_{-	0.02	}$ \\
24420	&	2000 BU22	&	11.45	&	21.723	$\pm$	1.211	&				&	1.65	$\pm$	0.7	&	0.55	$^{+	0.04	}_{-	0.04	}$ \\
24426	&	2000 CR12	&	12.13	&	14.336	$\pm$	1.007	&	0.71	$\pm$	0.054	&	4.64	$\pm$	0.8	&	0.49	$^{+	0.07	}_{-	0.07	}$ \\
28958	&	2001 CQ42	&	12.18	&	21.577	$\pm$	0.652	&				&	-0.04	$\pm$	0.8	&	0.41	$^{+	0.07	}_{-	0.07	}$ \\
39285	&	2001 BP75	&	12.49	&	17.602	$\pm$	0.499	&				&	0.25	$\pm$	0.69	&	0.48	$^{+	0.15	}_{-	0.13	}$ \\
39795	&	1997 SF28	&	12.42	&	18.342	$\pm$	0.742	&				&				&	0.54	$^{+	0.11	}_{-	0.09	}$ \\
43212	&	2000 AL113	&	12.2	&	19.212	$\pm$	1.09	&	0.56	$\pm$	0.028	&	1.19	$\pm$	0.78	&	0.53	$^{+	0.1	}_{-	0.09	}$ \\
43436	&	2000 YD42	&	12.12	&				&				&				&	0.45	$^{+	0.08	}_{-	0.08	}$ \\
53469	&	2000 AX8	&	12.39	&	18.453	$\pm$	0.354	&				&	0.17	$\pm$	0.8	&	0.44	$^{+	0.07	}_{-	0.06	}$ \\
65150	&	2002 CA126	&	12.47	&				&				&	4.14	$\pm$	0.7	&						\\
65225	&	2002 EK44	&	12.36	&	16.654	$\pm$	0.234	&	0.64	$\pm$	0.036	&	0.97	$\pm$	0.85	&	0.46	$^{+	0.1	}_{-	0.09	}$ \\
88229	&	2001 BZ54	&	12.32	&				&				&				&	0.39	$^{+	0.07	}_{-	0.07	}$ \\
111805	&	2002 CZ256	&	12.58	&				&	0.605	$\pm$	0.025	&				&	0.48	$^{+	0.14	}_{-	0.12	}$ \\
127846	&	2003 FO111	&	12.47	&				&				&				&	0.54	$^{+	0.1	}_{-	0.1	}$ \\
160856	&	2001 DU92	&	12.59	&	16.216	$\pm$	0.54	&				&				&	0.5	$^{+	0.07	}_{-	0.07	}$ \\
163135	&	2002 CT22	&	12.58	&	16.661	$\pm$	0.735	&				&	2.76	$\pm$	0.73	&						\\
163189	&	2002 EU6	&	12.9	&	16.23	$\pm$	0.781	&				&				&	0.53	$^{+	0.17	}_{-	0.14	}$ \\
163216	&	2002 EN68	&	12.55	&	13.25	$\pm$	0.801	&				&	3.6	$\pm$	0.98	&	0.52	$^{+	0.1	}_{-	0.09	}$ \\
166211	&	2002 EP135	&	12.88	&	14.412	$\pm$	1.052	&				&				&	0.57	$^{+	0.15	}_{-	0.14	}$ \\
191088	&	2002 CP286	&	12.95	&				&	0.8	$\pm$	0.072	&				&	0.5	$^{+	0.16	}_{-	0.14	}$ \\
192388	&	1996 RD29	&	12.91	&				&				&	2.76	$\pm$	0.89	&						\\
192929	&	2000 AT44	&	12.5	&	13.339	$\pm$	0.482	&				&	-0.53	$\pm$	0.83	&						\\
195412	&	2002 GF39	&	12.45	&	19.051	$\pm$	0.439	&				&				&	0.46	$^{+	0.08	}_{-	0.07	}$ \\
200024	&	2007 OO7	&	12.8	&	13.808	$\pm$	0.886	&				&				&	0.55	$^{+	0.16	}_{-	0.14	}$ \\
200032	&	2007 PU43	&	12.89	&	17.945	$\pm$	0.582	&				&				&	0.6	$^{+	0.15	}_{-	0.15	}$ \\
210237	&	2007 RQ154	&	12.75	&	16.698	$\pm$	0.713	&				&				&	0.5	$^{+	0.11	}_{-	0.09	}$ \\
214376	&	2005 LF20	&	13.06	&				&				&				&	0.52	$^{+	0.17	}_{-	0.15	}$ \\
219835	&	2002 CH82	&	13.29	&				&	0.72	$\pm$	0.05	&				&						\\
223251	&	2003 FB70	&	12.6	&	17.702	$\pm$	0.529	&	0.63	$\pm$	0.042	&				&						\\
237035	&	2008 SL91	&	13.37	&	11.96	$\pm$	0.773	&				&				&	0.59	$^{+	0.16	}_{-	0.14	}$ \\
246145	&	2007 PE9	&	13.21	&	12.093	$\pm$	0.601	&				&				&	0.46	$^{+	0.14	}_{-	0.12	}$ \\
252683	&	2002 AE166	&	13	&				&	0.58	$\pm$	0.028	&				&						\\
252711	&	2002 CU152	&	13.2	&				&	0.59	$\pm$	0.057	&				&						\\
259316	&	2003 FJ38	&	13.9	&				&	0.53	$\pm$	0.078	&				&						\\
313024	&	2000 AV210	&	13.6	&	10.805	$\pm$	0.599	&	0.96	$\pm$	0.092	&				&

\end{tabular}
\end{table}

\section*{Acknowledgments} 
\noindent The work in this paper was supported by the Lucy mission through NASA's Discovery Program grant NNM16AA08C.\\
R.M. and H.F.L. acknowledge the support from NASA's Emerging Worlds program, grant NNX17AE83G.\\
R.M. also acknowledges the support from the Swiss National Science Foundation (SNSF) under the grant P2BEP2\_184482.\\
D.N. acknowledges support from NASA SSW.\\
The work of R.D. was supported by the NASA's Emerging Worlds program, grant 80NSSC21K0387.\\
I.W. is supported by an appointment to the NASA Postdoctoral Program at the NASA Goddard Space Flight Center, administered by Oak Ridge Associated Universities under contract with NASA.

\bibliography{main}{}
\bibliographystyle{aasjournal}


\end{document}